\documentclass[a4paper,11pt]{article}
\pdfoutput=1 
\usepackage{jcappub}                     
\usepackage{appendix}
\usepackage[T1]{fontenc} 
\usepackage{graphicx}
\usepackage{subcaption}
\usepackage{graphicx}

\newcommand{\sgrA}{Sgr.~A$^\star$}

\title{\boldmath Boosted Dark Matter from Sagittarius A$^\star$}

\author[a,b]{Javier F. Acevedo,}
\author[a]{Adam Ritz}

\affiliation[a]{Department of Physics and Astronomy, University of Victoria, Victoria, BC V8P 5C2, Canada}
\affiliation[b]{TRIUMF, Vancouver, BC V6T 2A3, Canada}

\emailAdd{jfacev@uvic.ca}
\emailAdd{aritz@uvic.ca}

\abstract{It was recently demonstrated that black hole binaries can gravitationally accelerate ambient dark matter (DM), producing a continuous flux of particles with velocities far exceeding those of the galactic halo. 
We extend this analysis to the Milky Way’s nuclear star cluster, where stellar-mass black holes are expected to orbit in close proximity to the supermassive black hole Sagittarius A$^\star$. Using numerical simulations, we compute the flux of gravitationally-boosted DM sourced by this region. 
Because of the high DM density and large population of black holes orbiting deep within Sagittarius A$^\star$'s gravitational potential, the resulting DM ejecta attain substantially higher rates and energies compared to galactic black hole binaries, with simulated particles reaching velocities of up to $\sim 25,\!000 \, \rm km/s$. We find that the nuclear star cluster is therefore the dominant source of gravitationally-boosted DM in the Milky Way.
Even under conservative assumptions about the DM profile in the inner galaxy, the ejected DM flux from this region can render large-volume DM detectors competitive with lower-threshold experiments in the sub-GeV mass range, independently of the underlying DM particle model. The gravitational nature of the boost also opens up a sizable detection window into heavy inelastic DM scenarios that are otherwise largely inaccessible to conventional halo DM searches.}

\begin{document}
\maketitle
\flushbottom

\section{Introduction}
While noble liquid-based experiments are among the most sensitive probes available to test the particle nature of dark matter (DM), their reach using nuclear scattering events becomes limited when the DM mass is below the GeV scale. In this regime, the low mass of the DM, combined with its finite halo velocity of at most $\sim {\rm few} \times 10^{-3} \, c$, implies recoil energies that are too small to be efficiently detected by this class of detectors. Major experimental progress has been achieved through the development of low-threshold detectors sensitive to sub-GeV DM via collective excitations, molecular dissociation, and other low-energy channels, see $e.g.$ Ref.~\cite{Kahn:2021ttr}. These experiments are able to probe DM-nucleus scattering down to DM masses of $\sim 70$ MeV, albeit with a weaker cross-section sensitivity owing to their comparatively smaller size. This leaves a sizable window of otherwise viable DM-nucleus scattering parameter space beyond the reach of current direct detection experiments, particularly when compared to the same experimental constraints assuming DM-electron scattering.

As a complementary approach, a number of works have analyzed astrophysical processes capable of accelerating ambient halo DM, including cosmic-ray scattering \cite{Bringmann:2018cvk,Ema:2018bih,Alvey:2019zaa,Cappiello:2018hsu,Cappiello:2019qsw,Bell:2021xff,Maity:2022exk,Bell:2023sdq,Aitken:2026flk}, neutrino scattering \cite{Zhang:2020nis,Das:2021lcr,Lin:2022dbl}, supernovae \cite{Cappiello:2022exa}, solar reflection \cite{Kouvaris:2015nsa,Emken:2017hnp,An:2017ojc,Emken:2021lgc,An:2021qdl,Emken:2024nox} and blazar-boosting \cite{Granelli:2022ysi}. These mechanisms may up-scatter a small component of the galactic DM to higher energies, allowing DM and even neutrino detectors to probe mass values that are otherwise kinematically inaccessible. In these scenarios, both the flux and detectability of the boosted DM are controlled by the same non-gravitational couplings to the Standard Model (SM). Consequently, obtaining an observable flux requires relatively large couplings, while reducing these couplings to access new parameter space simultaneously suppresses the signal. Moreover, there is typically a finite DM mass range for which these boosting processes are effective. Related ideas have also been explored in specific particle-physics scenarios, where a boosted sub-population can arise from interactions within the dark sector itself \cite{Berger:2014sqa,Agashe:2014yua,Giudice:2017zke,Geller:2022gey,Acevedo:2024wmx}. While such approaches allow for a rich phenomenology, they are inherently tied to details of the underlying DM model.

In this context, binary systems have recently been shown to eject transiting halo DM particles to substantially higher energies through purely gravitational interactions \cite{Acevedo:2026xol}. In contrast with the previous processes, the gravitational nature of the boost implies that it is largely independent of the DM's specific particle properties. The energy gain can be understood from the so-called slingshot mechanism: in the limit where the DM mass is negligible, a close encounter with one component of the binary reduces to elastic gravitational scattering in that object's rest frame. In this frame, the DM particle approximately preserves its speed but is deflected in direction. When transforming the final velocity to the binary's center-of-mass frame, this deflection can increase the particle's energy. The scale of the energy gain is determined by the relative motion between the object that caused the deflection and the binary's center of mass, $i.e.$ its orbital velocity. Among the myriad of binary systems in the Milky Way, double black hole binaries are the most efficient for this process. Compared to any other stellar object, the typical black hole mass allows for fast moving orbits while maintaining large separations, enhancing both the rate and energy of DM ejections. Moreover, the nearly point-source gravitational field of black holes allows for strong deflections, permitting access to the full range of kinematically allowed ejection energies. Collectively, these systems accelerate a tiny sub-component of DM to higher velocities than its typical halo value, and can extend the mass reach of large-volume detectors to the sub-GeV regime for DM-nucleus scattering \cite{Acevedo:2026xol}.

In this work, we expand on this previous analysis and consider the boosted DM contribution from the nuclear star cluster, namely the concentration of gas, stars, and compact objects within a few parsecs from the supermassive black hole Sagittarius A$^\star$ (\sgrA) at the center of the Milky Way. Specifically, we focus on the deflection process by single, stellar-mass black holes directly orbiting about \sgrA, and compute the collective DM ejection rate from this population. Relative to galactic binaries comprised of dual stellar-mass black holes, this system entails a few crucial advantages, in part due to the extreme mass of \sgrA. First, a sizable number ($\sim 10^5$) of stellar-mass black holes is expected to reside within this region. Second, unlike galactic black hole binaries, these objects move with much larger speeds while maintaining a large orbital separation, which boosts both the rate and energy at which DM is ejected. Third, this region is widely understood to host the highest DM density in the Galaxy, potentially exceeding our local value by orders of magnitude. For these reasons, we find the nuclear star cluster to be the dominant galactic source of gravitationally-boosted DM. Based on the total flux of ejected DM, we assess the extended sensitivity to mass of large-volume direct detection experiments Lux-Zeplin (LZ), PandaX-4T, XENONnT and DarkSide-50, for benchmark spin-independent and spin-dependent elastic DM-nucleus scattering. In particular, we show that the boosted DM sourced by this region can render these experiments competitive with lower-threshold detectors at low masses in a nearly model-independent manner, even with conservative assumptions about the DM content around Sgr.~A$^\star$.

In addition, we consider inelastic spin-independent scattering whereby the initial and final DM states in a collision with a nucleus differ in mass. This imposes a kinematic threshold for scattering when the DM predominantly populates its lighter state, making detection challenging if the mass splitting between the DM states is sufficiently large, given the finite velocity of the particles in the halo. We focus here on the case where the lowest-lying DM state has a TeV-scale mass. As the boost is largely independent of mass, the flux from Sgr.~A$^\star$ enables detection of inelastic DM even at this scale, a unique feature of this mechanism compared to previous DM-boosting channels. We show how the contribution from the nuclear star cluster, assuming conservative DM density profiles, pushes the inelastic frontier of these detectors to inter-state mass splittings in the MeV range and virtually unexplored inelastic cross-sections.  

This paper is organized as follows: in Section~\ref{sec:ejection_analytic}, we first review a simple analytic description of the energy gain by DM through close gravitational interactions in binary systems, focusing on the case of extreme mass ratios. Section~\ref{sec:sim} introduces the Monte Carlo simulation along with applicable corrections for computing DM ejection rates deep in the gravitational potential of Sgr.~A$^\star$. In Section~\ref{sec:NSC_mod}, we discuss our modeling of the nuclear star cluster, both in terms of the distribution of orbital elements for the black holes populating this region as well as the assumed DM density and velocity dispersion. We compute the boosted DM flux and the associated velocity distribution in Section~\ref{sec:DM_flux}. In Section~\ref{sec:DD_prospects}, we evaluate the sensitivity reach of large-volume direct detection experiments for the aforementioned interaction benchmarks. We conclude in Section~\ref{sec:conclusions}. 

Throughout this work, we will use natural units where $\hbar = c = 1$ and $G = M^{-2}_{\rm planck}$. The energies of both the DM and orbiting black holes are normalized to mass. However, nuclear recoil energies in the context of direct detection are dimensionful.

\section{Analytic Description of Ejection}
\label{sec:ejection_analytic}
Before considering our numerical simulation below, we briefly review the gravitational ejection dynamics for a restricted 3-body problem, $i.e.$ the regime where $m_\chi \ll M_2 \ll M_1$ with $m_\chi$ being the DM mass and $M_2$ ($M_1$) the mass of the secondary (primary) binary component. As we are interested in binaries comprised of stellar-mass black holes orbiting around a supermassive black hole, this regime is of particular relevance here.  

\begin{figure*}[t!]
    \centering
    \hspace*{-0.9cm}
    \includegraphics[width=1.1\textwidth]{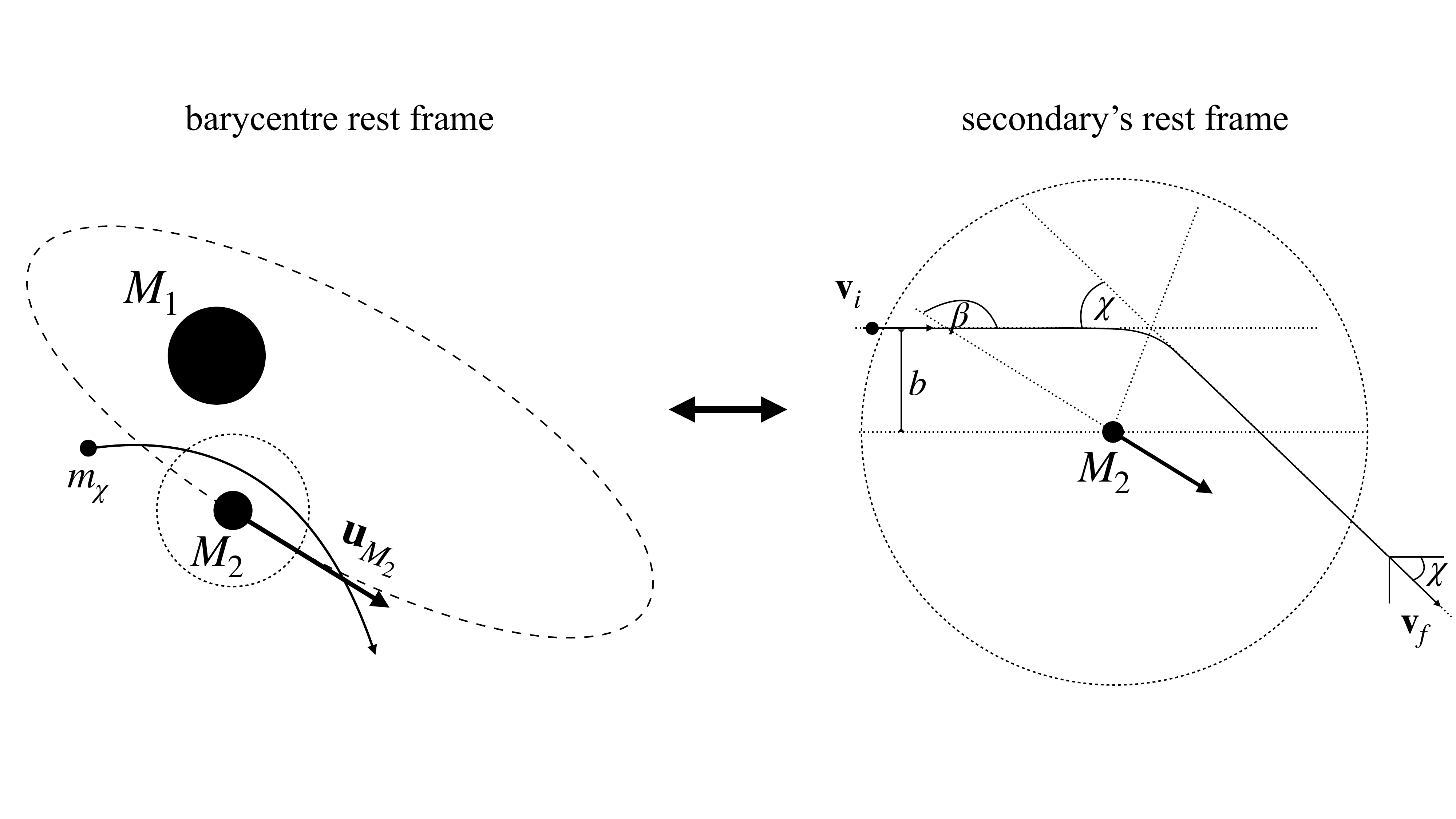}
    \caption{Deflection process for a DM particle interacting with a binary in the restricted regime $m_\chi \ll M_2 \ll M_1$, as viewed from the binary's barycenter frame (\textit{left}). Assuming the encounter is sufficiently close, and neglecting the energy lost or absorbed by the orbiting companion, the process can be described from the secondary's rest frame (\textit{right}), where the DM approaches with initial relative velocity $v_i$ and impact parameter $b$, and is deflected by an angle $\chi$. The angle $\beta$ parameterizes the approach relative to the secondary's velocity in the barycenter frame, which we draw explicitly on the right frame for clarity. In the barycenter frame, the DM particle gains or loses energy depending on whether its velocity projection onto the secondary's motion increases or decreases, respectively. See the text for further details.}
    \label{fig:slingshot_schem}
\end{figure*}

Figure~\ref{fig:slingshot_schem} illustrates the ejection process in both the binary's barycentre (center of mass) rest frame, as well as the secondary's rest frame. In the barycentre frame, the heavy primary mass $M_1$ is at rest, while the secondary mass $M_2$ is bound to it on a Keplerian orbit. Without loss of generality, we define this frame such that one of its axes is aligned with the secondary's velocity vector for this given point in the orbit. In the secondary's rest frame, an encounter between the DM and the orbiting secondary can be described by a simple two-body gravitational interaction with the scatterer at rest. This approximation is valid so long as the encounter is sufficiently close and short-lived. The DM enters the secondary's sphere of influence with (relative) velocity $\mathbf{v}_i$, and with an assumed impact parameter $b$. The former can be expanded in parallel and perpendicular components relative to the motion of the secondary, 
\begin{equation}
    \mathbf{v}_i = - v_{i}\,( \cos\beta, \, \sin\beta)~,
\end{equation}
where, $v_i$ is the relative velocity magnitude, and the angle $\beta$ parameterizes whether the DM predominantly approaches the secondary from its leading ($\beta = 0$) or trailing ($\beta = \pi$) side of motion. 
Neglecting the energy lost or absorbed by the secondary, the DM particle's velocity upon exiting the secondary's sphere of influence is simply the incoming velocity rotated by the scattering angle, $i.e.$ 
\begin{equation}
    \mathbf{v}_f = - v_{i}\,\left(\cos(\beta+\chi), \, \sin(\beta+\chi)\right)~,
\end{equation}
where $\chi$ is given by the standard formula for two-body gravitational scattering,
\begin{equation}
    \tan\left(\frac{\chi}{2}\right) = \frac{G M_2}{b \, v_i^2}~.
    \label{eq:scattering_angle}
\end{equation}
Upon boosting both initial and final velocity vectors back to the barycenter frame,
\begin{equation}
   \mathbf{v}_i + \mathbf{u}_{M_2} = \mathbf{u}_i = (u_{M_2} - v_{i}\cos\beta, -v_{i} \sin\beta)~,
\end{equation}
\begin{equation}
    \mathbf{v}_f + \mathbf{u}_{M_2} = \mathbf{u}_f = (u_{M_2} - v_{i}\cos(\beta+\chi) , -v_{i} \sin(\beta+\chi))~,
\end{equation}
the deflection process has now shifted the DM particle's energy in this frame by an amount
\begin{equation}
    \Delta \varepsilon = \frac{1}{2} \left(u_f^2 - u_i^2\right) = u_{M_2} v_{i} \, \left(\cos\beta - \cos(\beta+\chi)\right)~.
    \label{eq:delta_epsilon}
\end{equation}
Thus, at the lowest order, the energy gain can be understood from a simple transformation between frames, without the need to model in detail the energy loss or absorption by the binary itself. In fact, any energy exchanged will scale as $m_\chi / M_{2}$, and therefore will be completely negligible given the DM mass range we focus on.

Eq.~\eqref{eq:delta_epsilon} indicates that the energy loss or gain depends on whether the DM particle increased its velocity projection onto the direction of motion of the secondary, and the scale of the energy change is proportional to the secondary's orbital speed $u_{M_{2}}$. For instance, when $\beta < \pi/2$, $\Delta \varepsilon > 0$ for any scattering angle $\chi$, because any deflection will always increase the projected velocity of the DM onto $\mathbf{u}_{M_{2}}$. If $\beta > \pi/2$, by contrast, only some range of scattering angles will increase the projected velocity onto $\mathbf{u}_{M_{2}}$. The maximal energy gain naturally occurs when $\beta = 0$ and $\chi = \pi$, corresponding to a particle approaching the secondary from its leading side of motion, and being fully deflected in the same direction as the secondary, gaining an amount $\Delta \varepsilon \simeq 2 \, v_i \, u_{M_{2}}$. We remark that the energy change, within the approximations made here, only depends on the mass of the secondary through the scattering angle $\chi$. Consequently, two orbiting companions of different masses but otherwise in the same orbital configuration share the same kinematically allowed range of ejection energies. The difference lies in that, for a given ejection energy, the heavier object will eject DM more frequently compared to the lighter object, since larger impact parameters achieve the same deflection, allowing in turn for a larger fraction of the incoming flux to achieve the required scattering angle. We provide additional details below of how the ejection rate scales with the mass of the secondary. 

We emphasize that this treatment mainly applies for compact objects, which can largely be treated as point-like masses for computing the gravitational scattering of DM. Extended systems, such as stars or planets, cannot be modeled in this way for sufficiently close encounters. Finite size effects must be taken into account in this case, both through deviations from the Newtonian $1/r$ potential and, depending on the model, the possibility of DM scattering within the object's interior. For example, for the same orbital parameters, an orbiting main-sequence star will not achieve the same ejection energies as a black hole. In the case of the star, the DM impact parameter is bounded from below by its physical radius (or some other similar scale), preventing the strong deflections required for ejections near the maximum energy. A more detailed discussion of binaries with objects other than black holes can be found in Ref.~\cite{Acevedo:2026xol}.

Within our specific context, Eq.~\eqref{eq:delta_epsilon} shows that black holes on short-period and/or highly-eccentric orbits around Sgr.~A$^\star$ will be capable of producing a hardened spectrum of ejections, as the DM particles are deflected by rapidly moving scatterers deep within the nuclear star cluster’s potential. For instance, given our binning (discussed in detail below) of the nuclear star cluster's phase space, the most energetic ejections correspond to a semimajor axis $a \simeq 386 \ \rm AU$ and eccentricity $e \simeq 0.94$. Given the scaling of Eq.~\eqref{eq:delta_epsilon} with orbital velocity, the maximum energy gain must occur during pericenter passage. Ignoring relativistic corrections, which are mild in this regime, the pericenter velocity of the orbiting black hole is $u_{M_2} \simeq 1.79 \times 10^{4} \ \rm km/s$. Assuming a head-on encounter ($i.e.$ $\beta = 0$) exactly at pericenter, the relative velocity is the sum of the black hole's pericenter velocity, and the DM's velocity in the barycenter frame, the latter being approximately given by \sgrA's escape velocity. For a pericenter distance of $\sim 23 \ \rm AU$, this is $\sim 1.82 \times 10^4 \ \rm km/s$. Assuming a small enough impact parameter such that $\chi \simeq \pi$, the resulting maximum energy is $\Delta \varepsilon \simeq 1.4 \times 10^{-2}$, translating into a semi-relativistic ejection speed of $\sim 5.1 \times 10^4 \ \rm km/s$. In practice, however, we numerically observe ejecta with velocities less than half this maximum value, given the vanishingly low probability of such pericenter ejection events. 

\section{Simulation of Gravitational Ejection}
\label{sec:sim}
We employ the Monte Carlo simulation introduced in Ref.~\cite{Acevedo:2026xol}, which evolves in parallel a large ensemble of 3-body systems comprised of two binary objects and a third DM test particle, using a Velocity Verlet algorithm \cite{hairer2006geometric} in the binary's barycentre rest frame. 

In each realization, the two heavy masses are initialized with positions and velocities corresponding to a Keplerian orbit, with the true anomaly sampled uniformly over the orbit. The DM is injected on a boundary sphere, with a uniform distribution in position and a randomized velocity sampled from a specified distribution. A cutoff is applied to the velocity magnitude such that only \textit{unbound} DM particles are simulated. While simulating bound DM particles is also admissible under this scheme, as discussed further below, the interpretation of the simulation output in this case requires more caution. The system is then evolved until the DM is either captured or ejected by the binary, though in practice nearly all realizations result in rapid ejection. From the ensemble, we construct the differential DM ejection rate as a function of energy and direction. This method carries almost no assumptions concerning the particle nature of DM, its only limitations being: 1) the DM must not be too light or too heavy, in order to respectively avoid the wave-like regime and back-reaction on the binary, and 2) any DM self-interactions must be sufficiently weak for the evolution of the DM particles to be solely determined by the binary's gravitational potential. 

The main simulation output is the differential ejection rate, which we compute as \cite{Acevedo:2026xol}
\begin{equation}
   m_\chi \, \frac{d^3N_\chi}{d\Omega \, d\varepsilon \, dt} \simeq \mathcal{F}_{\rm sim} \times \frac{d^2p}{d\Omega \, d\varepsilon}~.
    \label{eq:diff_spec_gen}
\end{equation}
The right-hand side factorizes into an incoming flux factor $\mathcal F_{\rm sim}$, and an ejection probability distribution $d^2p/d\Omega\,d\varepsilon$.

The flux factor incorporates the physical rate at which unbound DM particles would enter the simulation volume, given the input velocity distribution for the DM particles far from the binary system, which we assume to be Maxwellian. In Appendix~\ref{app:sim_factor}, we show that this factor is approximately given by
\begin{equation}
    \mathcal{F}_{\rm sim} = \sqrt{\frac{2}{\pi}} \, 4 \pi R_{\rm sim}^2 \, \rho^{\rm sim}_\chi \, \sigma^{\rm sim}_\chi
    \label{eq:Fsim_gen}
\end{equation}
where 
\begin{equation}
    \rho^{\rm sim}_\chi = \rho_\chi \,\sqrt{1+\frac{|\phi_{\rm sim}|}{\sigma^2_\chi}}~,
    \label{eq:rho_chi_sim}
\end{equation}
\begin{equation}
    \sigma^{\rm sim}_\chi = \sigma_\chi \,\sqrt{1+\frac{|\phi_{\rm sim}|}{\sigma^2_\chi}}~.
    \label{eq:sigma_chi_sim}
\end{equation}
Eq.~\eqref{eq:Fsim_gen} is the expression for the flux through a sphere of radius $R_{\rm sim}$, given a density $\rho^{\rm sim}_\chi$ and velocity dispersion $\sigma^{\rm sim}_\chi$ at the boundary. These are in turn given by asymptotic values $\rho_\chi$ and $\sigma_\chi$ far from Sgr.~A$^\star$'s influence, amplified by its gravitational potential $|\phi_{\rm sim}|$ at the boundary. In Sec.~\ref{sec:NSC_mod}, we specify and justify our choice for $\rho_\chi$ and $\sigma_\chi$. The inclusion of the gravitational focusing factors is necessary in all of our runs, as the simulation boundary lies sufficiently close to Sgr.~A$^\star$ for the background density and velocity to be distorted by its presence. 

The second factor in Eq.~\eqref{eq:diff_spec_gen} represents the probability of a particle being ejected in a direction given by the solid angle $\Omega$ and final energy per unit mass $\varepsilon$. While these two variables are not strictly independent, it was shown in Ref.~\cite{Acevedo:2026xol} that the ejection direction can be reasonably approximated as isotropic and independent of energy. In other words, we approximate
\begin{equation}
    \frac{d^2p}{d\Omega \, d\varepsilon} \simeq \frac{1}{4 \pi} \frac{dp}{d\varepsilon}~,
    \label{eq:iso_approx}
\end{equation}
where $dp/d\varepsilon$ measures the probability of ejection with energy $\varepsilon$ marginalized along all directions. 

We emphasize that Eq.~\eqref{eq:diff_spec_gen} produces a two-component spectrum: a low-energy component from weak, distant scatterings and a high-energy component from close encounters with the binary. The low-energy component ultimately originates from the simulation recording infinitesimal energy changes in an $\mathcal{O}(1)$ fraction of all simulated particles. In fact, its normalization depends on the simulation boundary: all injected particles, up to an impact parameter set by the simulation size, undergo soft gravitational scatterings. This unphysical behavior can for instance be eliminated by imposing a finite energy gain threshold for the ejected particle to be counted by the simulation, which is equivalent to imposing some maximum impact parameter that does not trace the simulation boundary size (for further details, see Ref.~\cite{Acevedo:2026xol}). Throughout this work, we focus on the high-energy tail of the ejection spectrum, which converges to the same value regardless of the assumed simulation volume. This component approximately begins at the velocity cutoff of the underlying halo distribution, since DM particles do not naturally populate velocities much above this scale unless they have undergone a hard gravitational encounter.

The extreme mass ratio regime we focus on also allows us to simplify some aspects of the simulation without sacrificing precision. First, we fix Sgr.~A$^\star$ at the origin, rendering this particle effectively static and therefore decreasing the number of position and velocity evaluations in each timestep. Comparing the output against the full simulation in which Sgr.~A$^\star$ is allowed to dynamically evolve, we have confirmed that this approximation has no impact on the resulting ejection spectrum. Second, convergence of the ejection spectrum can be accelerated by simulating an artificially heavier secondary, provided the system remains in the extreme mass ratio regime. As noted in the previous section, a heavier secondary deflects particles more frequently than a lighter one in the same orbital configuration, while maintaining the same accessible range of ejection energies. In fact, in the extreme mass ratio regime, the physical component of the numerical ejection spectrum globally scales as $M_2^2$, $i.e.$
\begin{equation}
    \frac{d^2p}{d\Omega \, d\varepsilon}\bigg|_{M_2} = \left(\frac{M_2}{M^\prime_2}\right)^2 \, \frac{d^2p}{d\Omega \, d\varepsilon}\bigg|_{M^\prime_2}~,
\end{equation}
where $M_2$, $M^\prime_2$ denote different simulated masses.
This scaling directly originates from the standard gravitational cross-section, and was shown to hold to high accuracy in Ref.~\cite{Acevedo:2026xol}. Therefore, simulating a more massive secondary reduces the number of particles required to achieve the desired level of convergence. The resulting spectrum is then rescaled to the target secondary mass.

Finally, we remark that our simulation only assumes Newtonian gravity. We find this to be a reasonable approximation, since the smallest simulated pericenter distance for the orbiting black hole is $\sim 23 \ \rm AU$, and therefore sufficiently large for any post-Newtonian corrections to be negligible for the purposes of this work. Additionally, in each individual 3-body realization, a DM particle is removed from the sample if at any time it approaches either black hole to within a distance $< 10 \times$ its gravitational radius. In practice, such events only constitute a negligible fraction of the full simulated ensemble. 

\section{Modeling the Nuclear Star Cluster}
\label{sec:NSC_mod}
The differential ejection spectrum, as defined by Eq.~\eqref{eq:diff_spec_gen}, requires as inputs both the orbital elements of the black hole, as well as the underlying DM density and velocity dispersion. In what follows, we review our modeling of how these quantities are distributed within the central parsec surrounding Sgr.~A$^\star$.

\subsection{Black Hole Distribution}
We will model the nuclear star cluster population as being in both a spherically-symmetric and steady-state configuration. These assumptions are justified given the relatively orbital small distances we focus on compared to the nuclear star cluster size; at most of order $\sim 0.2 \ \rm pc$. At these distances, the potential is dominated by Sgr.~A$^\star$'s mass, and the two-body relaxation time is generically expected to be much shorter than the nuclear star cluster's age. Further discussion of the relaxation timescale can be found in Appendix~\ref{app:timescales}. 

By Jeans theorem, the above assumptions imply that the phase space distribution of the black hole population is a function of specific energy and the magnitude of angular momentum. However, for simplicity, we will drop the angular momentum dependence, which corresponds to assuming an isotropic velocity field. This is in fact justified by the frequent two-body interactions within such a small volume, which should produce a near-thermal distribution for the orbital elements over the age of the nuclear star cluster. 

Motivated by existing analytic solutions in this isotropic, steady-state approximation, we parameterize the distribution function as a power-law of the form,
\begin{equation}
    \frac{d^2N_{\rm BH}}{d \mathbf{r} \, d\mathbf{v}} = f(\varepsilon) \propto \varepsilon^p~,
    \label{eq:DF_main}
\end{equation}
where we specify the slope $p$ below, and omit for the moment the normalizing factor. The specific energy $\varepsilon$ is given by 
\begin{equation}
    \varepsilon = - \frac{v^2_r}{2} - \frac{J^2}{2 r^2} + \phi(r) = \frac{GM_1}{2 a}~,
    \label{eq:specific_E}
\end{equation}
where $r$ is the radial distance, $\phi(r)$ is the gravitational potential of the nuclear star cluster, $v_r$ is the radial velocity, and $J$ is the specific angular momentum,
\begin{equation}
    J = |\mathbf{r} \times \mathbf{v}| = \sqrt{GM_1 \, a \,(1-e^2)}~.
\end{equation}
We remark that we have defined the specific energy with a minus sign relative to its usual definition, so that $\varepsilon > 0$ corresponds to bound states. The rightmost expressions for $\varepsilon$ and $J$ are written in terms of semimajor axis $a$ and eccentricity $e$ assuming 
\begin{equation}
    \phi(r) = \frac{G M_1}{r}~,
\end{equation}
where $M_1 \simeq 4 \times 10^6 \, M_\odot$ is Sgr.~A$^\star$'s mass \cite{2016ApJ...830...17B,2017ApJ...837...30G,EventHorizonTelescope:2022wkp,EventHorizonTelescope:2022exc}. In other words, we approximate the orbits as Keplerian under Sgr.~A$^\star$'s long-distance gravitational field, which is appropriate given the smallest pericenter distances we simulate are enormous compared to its gravitational radius. 

From Eq.~\eqref{eq:DF_main}, as well as the subsequent definitions of energy and angular momentum, the joint semimajor axis and eccentricity distribution can be straightforwardly derived (see $e.g.$ Ref.~\cite{Schodel:2003gy})
\begin{equation}
    \frac{d^2N_{\rm BH}}{da \, de} \propto e \, a^{1/2 - p} ~.
    \label{eq:dist_sma_eccec}
\end{equation}
Thus, in terms of the orbital elements, the resulting distribution also scales as a double power-law in semimajor axis and eccentricity $e$. Note that these two variables are not strictly independent; this is because in general there is a maximum value of $J$ such that $\varepsilon > 0$. For our purpose of binning over the black hole population, it is convenient to reformulate Eq.~\eqref{eq:dist_sma_eccec} in alternative variables. We first convert Eq.~\eqref{eq:dist_sma_eccec} to a function of pericenter $r_p = a (1-e)$ and apocenter $r_a = a (1+e)$ distances. The corresponding transformation Jacobian introduces an additional factor $(r_a + r_p)^{-1}$, yielding 
\begin{equation}
    \frac{d^2N_{\rm BH}}{dr_p \, dr_a} \propto (r_a - r_p) (r_a + r_p)^{-3/2 -p}~.
\end{equation}
This distribution is defined in a triangular domain set by the conditions $0 \leq r_p \leq r_a$ and $r_a \leq R_{\rm NSC}$. The last inequality, in particular, enforces the orbits to be fully contained within our analysis region set by a maximum distance $R_{\rm NSC}$, which we specify below. We now introduce new variables 
\begin{equation}
    r_{\pm} = r_a \pm r_p~.
    \label{eq:r_plus_minus}
\end{equation}
We omit the Jacobian in this case, as it is a constant that can be absorbed in the normalization. We then obtain
\begin{equation}
    \frac{d^2N_{\rm BH}}{dr_{+} \,dr_{-}} \propto r_{-} \, r_{+}^{-3/2-p}~,
    \label{eq:dist_rplusminus}
\end{equation}
where the domain is now $r_{-}\leq r_{+} \leq 2R_{\rm NSC} - r_{-}$ and $0 \leq r_{-} \leq R_{\rm NSC}$. Note that, in particular, the $r_{-} = r_{+}$ boundary corresponds to linear orbits, whereas the $r_{-} = 0$ boundary corresponds to circular ones. The remaining inequalities require that, as before, the orbits are fully contained within the analysis region given by the distance $R_{\rm NSC}$.  

\begin{figure}[h]
\centering
\vspace{-0.9cm}
\begin{subfigure}{0.47\textwidth}
    \includegraphics[width=\linewidth]{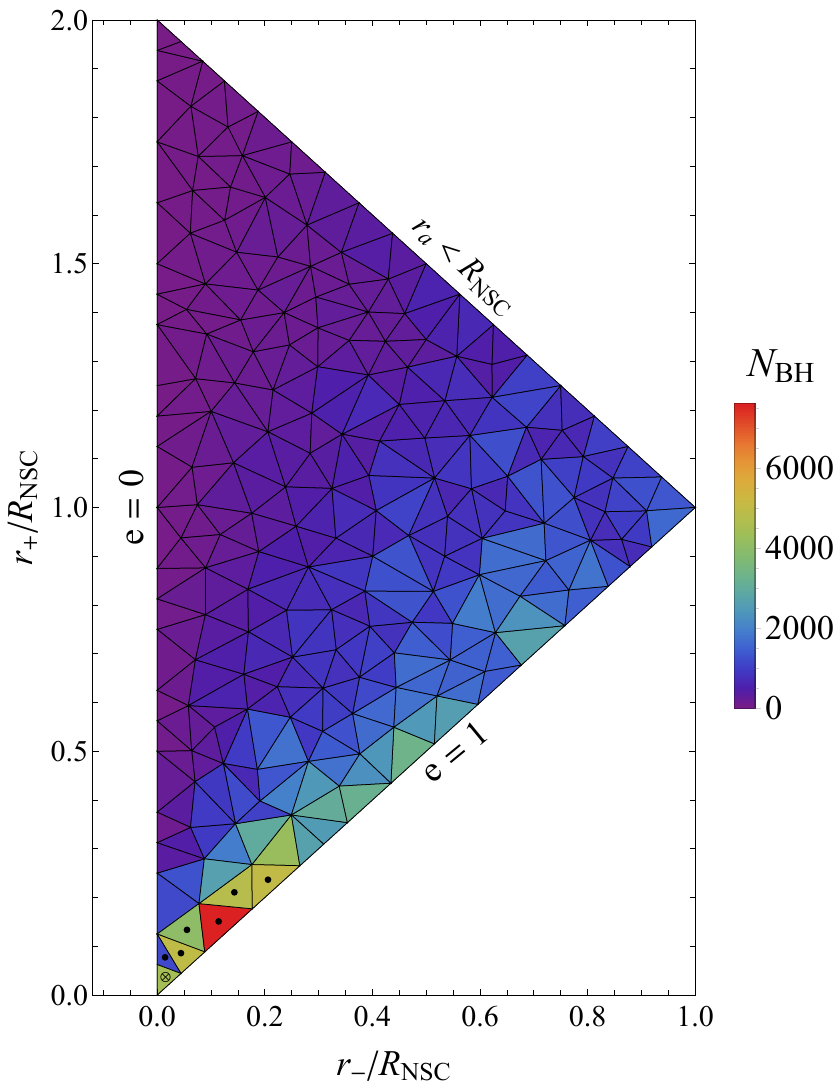}
\end{subfigure}
\hfill
\begin{subfigure}{0.456\textwidth}
    \includegraphics[width=\linewidth]{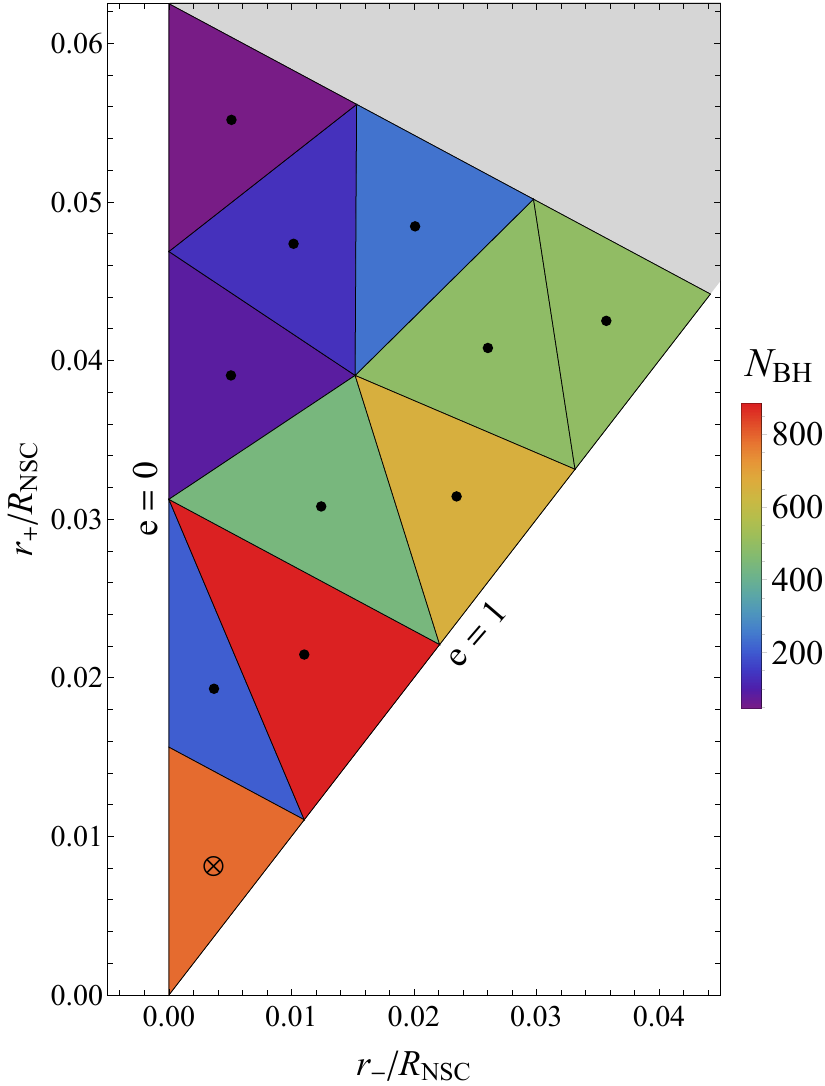}
\end{subfigure}
\vspace{0.62cm}
\hspace{-0.51cm}
\begin{subfigure}{0.472\textwidth}
    \includegraphics[width=\linewidth]{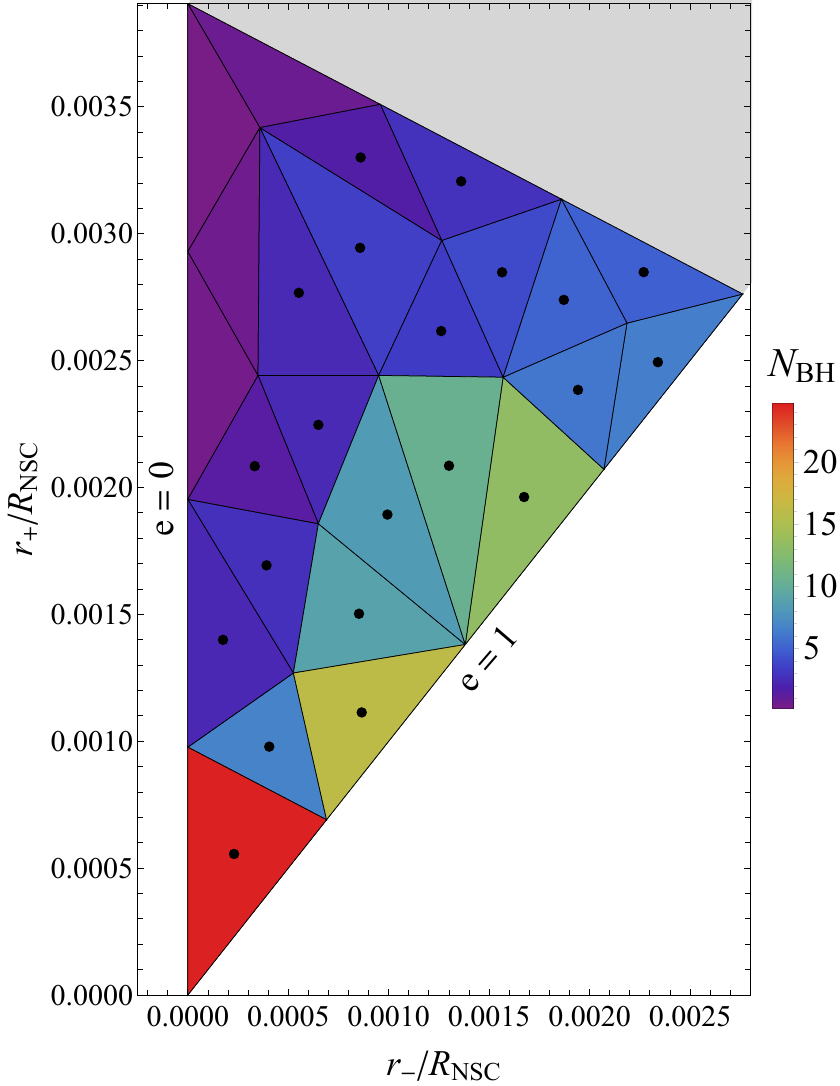}
\end{subfigure}
\hfill
\begin{subfigure}{0.472\textwidth}
    \includegraphics[width=\linewidth]{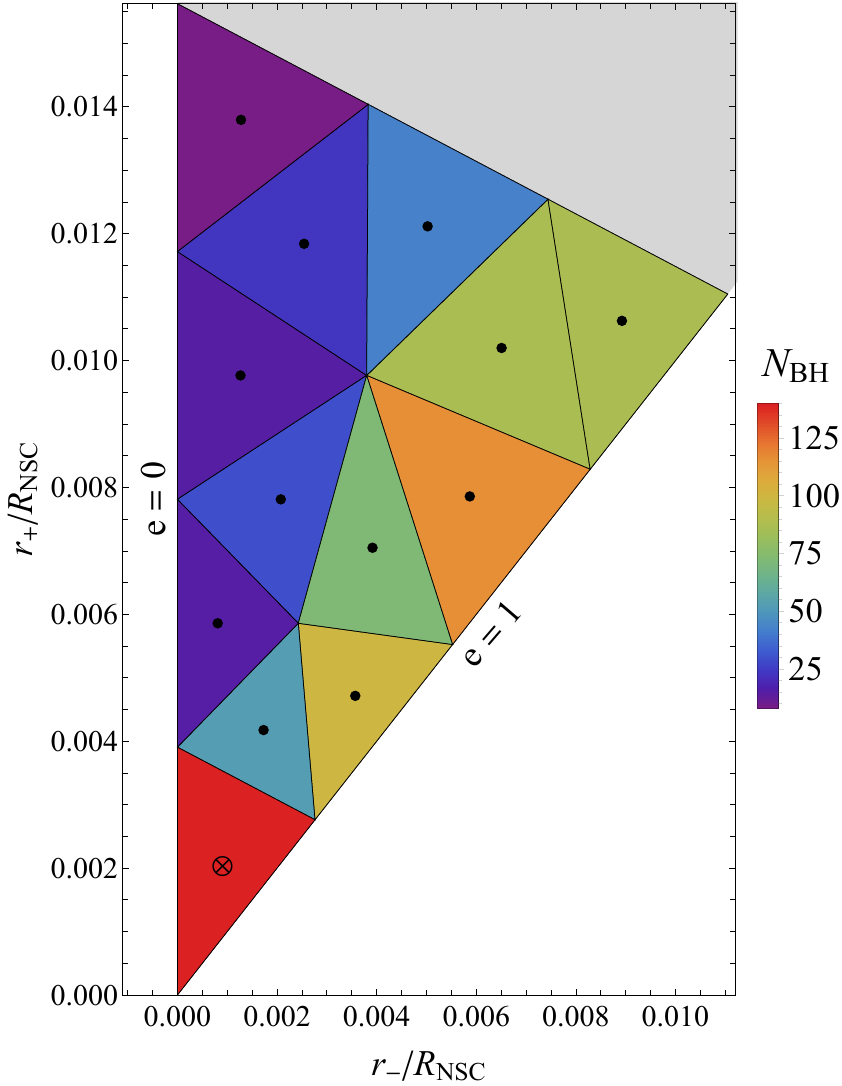}
\end{subfigure}
\caption{Simulation grid for the black hole population assuming a Bahcall-Wolf cusp profile, $cf.$ Eq.~\eqref{eq:dist_rplusminus}. The boundaries indicate the eccentricity range and maximum apocenter distance. The colour scale shows the expected number of black holes per bin. Dots mark the simulated bins and their associated triangular centroids, from which we extract the simulation parameters. Bins marked with a cross are sub-divided to resolve the high-energy tail of the ejecta; when read clockwise, each panel shows a successive iteration of this zoom-in process.}
\label{fig:sim_grid_BW}
\end{figure}

The slope $p$ in Eq.~\eqref{eq:dist_rplusminus} has a relatively wide uncertainty, spanning from $p \simeq 0.22$ up to $p \simeq 1.25$, based on reported values from both theoretical \cite{1977ApJ...216..883B,Quinlan:1994ed,1980ApJ...242.1232Y,Hopman:2006xn,Alexander:2008tq} and simulation works \cite{Amaro-Seoane:2010dzj,Panamarev:2018bwq,2018A&A...609A..28B,2024ApJ...961..232Z}. Distribution functions of the above form are also supported by observations of stars \cite{2018A&A...609A..26G,2019ApJ...872L..15H} and x-ray binaries \cite{2018Natur.556...70H} in this region. As our benchmark, we will use $p = 1/4$, corresponding to the seminal Bahcall-Wolf cusp \cite{1977ApJ...216..883B} in the limit that the stellar-mass black holes are the heaviest object across the mass spectrum of the nuclear star cluster. This is in fact a conservative choice for the scope of this work: the shallow slope implies short-period orbits will be less populated relative to more aggressive choices of this parameter, therefore suppressing the flux of boosted DM at the highest energies, and in turn limiting the mass reach of large-volume detectors. The overall normalization of Eq.~\eqref{eq:dist_rplusminus} will be set by the total number of black holes contained within the nuclear star cluster $N_{\rm BH}$,
\begin{equation}
    \int^{R_{\rm NSC}}_0 dr_{-} \int^{2R_{\rm NSC} - r_{-}}_{r_{-}} dr_{+} \, \frac{d^2N_{\rm BH}}{dr_{+} dr_{-}} = N_{\rm BH}
\end{equation}
This integral is convergent throughout the full range of slopes $p$ discussed above. To determine both $N_{\rm BH}$ and $R_{\rm NSC}$, we consider the most recent analysis of the star formation history of the nuclear star cluster, which indicates $N_{\rm BH}\simeq 2.5 \times 10^5$ single black holes contained within $R_{\rm NSC} \simeq 1.5 \ \rm{pc}$ of Sgr.~A$^\star$ \cite{2023ApJ...944...79C}. These estimates exceed earlier predictions \cite{Hopman:2006xn,Alexander:2008tq}, largely due to the inclusion of more recent observations indicating high metallicity in this region \cite{2015A&A...573A..14R,2015ApJ...809..143D}. We will mostly assume a peaked mass spectrum centered at $10 \, M_\odot$ for the orbiting black holes, which is the standard choice for numerical simulations in this region, although below we also analyze possible degeneracies between the mass spectrum and black hole number. 

Figure~\ref{fig:sim_grid_BW} shows our adopted simulation grid in the $(r_+, r_-)$ plane. The triangles in each panel denote bins in orbital element space, constructed from Mathematica’s built-in \texttt{ToElementMesh} function. The dots indicate the triangular centroid, from which we extract the simulation parameters for each bin via Eq.~\eqref{eq:r_plus_minus}. The color scale indicates the expected number of black holes per bin, which are distributed per Eq.~\eqref{eq:dist_rplusminus}. The upper left panel shows the initial discretization of the domain. The remaining panels, when read clockwise, show further sub-divisions of the lowest-period bin, marked by a cross in each case. This refinement is necessary to properly resolve the highest energy ejections, as these are produced by the black holes that orbit the closest to Sgr.~A$^\star$. The zoom-in procedure on the lowest-period orbital bin is terminated for an $a \simeq 86 \ \rm AU$, $e \simeq 0.41$ orbit, populated by $N_{\rm BH} \simeq 24$ black holes.  We find that further sub-divisions result in a mass and cross-section reach for DM experiments which only differs by an $\mathcal{O}(1)$ factor, which is sufficient given the scope of this work. Bins without a dot in the top- and bottom-left panels are excluded from our analysis. In the top-left panel, these correspond to black holes on extended orbits which cannot produce ejecta significantly above the typical halo velocity. In the bottom-left panel, bins with low eccentricities are discarded if they are expected to host fewer than one black hole. The total number of black holes populating the simulated bins is $\sim 3.2 \times 10^4$, and therefore a small fraction of the total number assumed to reside within the nuclear star cluster.

Finally, we remark that there exist limits on the extended dark mass surrounding Sgr.~A$^\star$ based on the prograde precession of the star S2, which corresponds to $\lesssim 4300 \, M_\odot$ at $3\sigma$ for a power-law slope matching that of our analysis \cite{GRAVITY:2024tth}. However, this bound applies to a diffuse, spherically symmetric mass distribution and can be significantly relaxed if the mass is instead granular in nature \cite{Bordoni:2025mli}, as would be the case for a population of stellar-mass black holes.
Even setting aside this caveat and taking the bound at face value, one can use Eq.~\eqref{eq:DF_main} to compute the associated black hole density (see $e.g.$ Ref.~\cite{Schodel:2003gy}) which, for an arbitrary slope $p$, scales as 
\begin{equation}
    \frac{dN_{\rm BH}}{d\mathbf{r}} \propto  r^{-3/2-p}~,
    \label{eq:BH_dens_cusp}
\end{equation}
where once again we have omitted the normalization constant for simplicity. 
For our fiducial choice of black hole number and mass, the above approach yields $\sim 4440 \, M_\odot$ within S2's apocenter distance, remaining approximately consistent with the observational constraint at this confidence level. Moreover, because the ejection spectrum scales quadratically with black hole mass, a smaller population of more massive black holes can produce a comparable or even larger boosted dark matter flux relative to a more numerous population of lighter black holes. As we illustrate below, there are combinations of black hole number and mass that yield a comparable signal while remaining significantly below this putative enclosed mass limit.

\subsection{Dark Matter Density and Velocity}
On the scale of the Milky Way, we assume a Navarro-Frenk-White (NFW) DM halo density profile \cite{Navarro:1995iw,Navarro:1996gj},
\begin{equation}
    \rho_{\chi}(r)  = \dfrac{\rho_{s}}{\left({r}/{r_s}\right)^{\gamma} \left(1+{r}/{r_s}\right)^{3-\gamma}}~,
    \label{eq:NFW_profile}
\end{equation}
where $r_s \simeq 20 \ \rm kpc$ is the scale radius, and $\gamma$ is its slope. The parameter $\rho_{s}$ is fixed so that $\rho_\chi \simeq 0.42 \ \rm GeV \ cm^{-3}$ at the Sun's position \cite{Pato:2015dua}. For the slope, we consider a possible range from $\gamma = 1$ up to $1.5$ based on halo contraction analyses of the inner Galaxy~\cite{2011arXiv1108.5736G,DiCintio:2014xia}. The DM's velocity dispersion profile is not well-known towards the galactic center. However, observations indicate a relatively weak variation within the central $\sim 100 \ \rm pc$, staying within the $\sim 200 - 300 \ \rm km/s$ range typically assumed for our local position in the galaxy \cite{2013PASJ...65..118S}. Therefore, for simplicity, we assume a constant $\sigma_\chi(r) = 240 \ \rm km/s$ dispersion profile.

Both the density and velocity dispersion profiles determine the asymptotic values for these quantities outside the sphere of influence of Sgr.~A$^\star$ at some matching radius. Below this scale, the unbound DM density and velocity dispersion are extrapolated using Eqs.~\eqref{eq:rho_chi_sim} and \eqref{eq:sigma_chi_sim} (see also App.~\ref{app:sim_factor}). Conversely, beyond this scale, the majority of DM particles are expected to be unbound to Sgr.~A$^\star$, with the halo acting as a reservoir that continuously supplies particles passing through the nuclear star cluster. To determine the matching radius, we consider the galactocentric distance at which the escape velocity of Sgr.~A$^\star$ equates to the assumed velocity dispersion ($i.e.$ $|\phi(r)| \simeq \sigma_\chi^2$), which yields $\sim 1.5 \ \rm pc$. This procedure ensures that the gravitational focusing induced by Sgr.~A$^\star$ is applied exclusively within the region where its potential dominates, while at larger radii the density and velocity dispersion remain anchored to the galactic-level profiles.

By restricting ourselves to the unbound particles traversing the nuclear star cluster, we are likely underestimating the full DM content in this region. For instance, assuming a standard $\gamma = 1$ NFW profile, the integrated density as given by Eq.~\eqref{eq:rho_chi_sim} predicts an enclosed mass $\sim 6 \times 10^{-3} \, M_\odot$ within S2's apocenter. This is negligible compared to the value obtained by extrapolating the full NFW profile, which gives $\sim 0.15 \, M_\odot$ enclosed within the same volume.
Incorporating also the bound DM population around Sgr.~A$^\star$ would, however, require a self-consistent treatment of depletion over time due to repeated ejections, which lies beyond the ensemble-based approach of our simulation. This nevertheless highlights our conservative approach to the DM profile, as we are only considering a tiny subpopulation predicted by standard halo profiles without the need to invoke, for example, DM spikes.

\section{Boosted DM Velocity Distribution}
\label{sec:DM_flux}
Having computed the approximate ejection spectrum for the black hole population bound to Sgr.~A$^\star$, we now proceed to compute the boosted DM flux, and the associated velocity distribution to be used for estimating direct detection prospects. 

We first construct the integrated ejection spectrum of the nuclear star cluster. This is obtained by stacking the individual contributions from each orbital element bin, identified by a unique pair of semimajor axis $a$ and eccentricity $e$ values,
\begin{equation}
    \left(\frac{d^2N_\chi}{dt\, d\Omega \, d\varepsilon}\right)_{\rm NSC} = \frac{1}{4\pi} \sum_{a,e} N_{a,e} \, \mathcal{F}^{(a,e)}_{\rm sim} \, \frac{dp}{d\varepsilon}\bigg{|}_{a,e}~.
    \label{eq:integrated_spec}
\end{equation}
The coefficients $N_{a,e}$ are the expected number of black holes populating each bin. The other two factors in each term are given by Eqs.~\eqref{eq:diff_spec_gen} and \eqref{eq:iso_approx}. We have explicitly added summation indices to both the flux factor $\mathcal{F}_{\rm sim}$ and the ejection energy distribution $dp/d\varepsilon$, as these are different for each combination of orbital elements. 

\begin{figure*}[t!]
    \centering
    \includegraphics[width=0.9\textwidth]{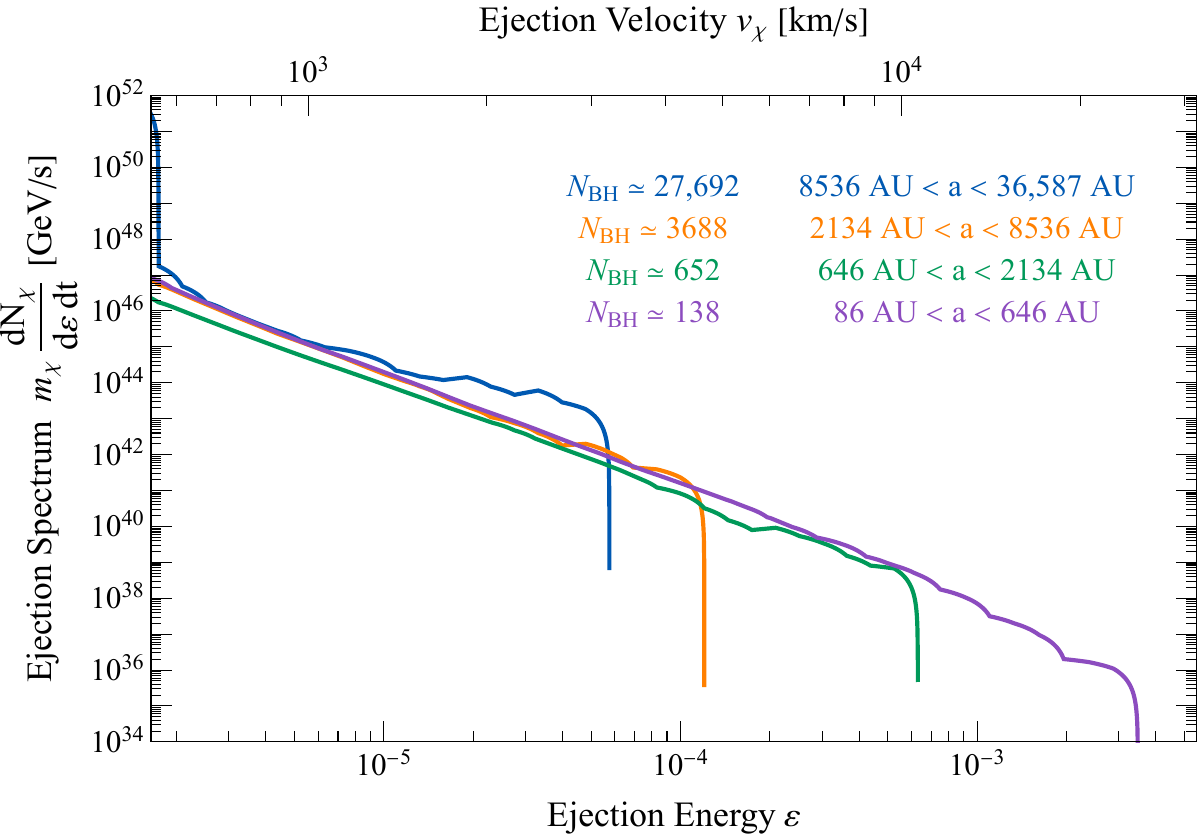}
    \caption{Numerical ejection spectra for each panel of Fig.~\ref{fig:sim_grid_BW}, assuming $10 \, M_\odot$ for the black holes and a standard NFW profile with slope $\gamma = 1$ at the boundary of the nuclear star cluster. The total black hole number and semimajor axis range spanned by the simulated bins within each grid are indicated by color. Note that the binning approach adopted to sample the black hole distribution leads to some numerical noise, which is visible as kinks in the output spectrum.}
    \label{fig:Bin_Spectra}
\end{figure*}

Figure~\ref{fig:Bin_Spectra} shows the ejection spectra obtained from Eq.~\eqref{eq:integrated_spec} considering all bins within each orbital element grid, with the four resulting spectra corresponding to the four panels of Fig.~\ref{fig:sim_grid_BW}. The boundary DM density is assumed to follow an NFW profile with slope $\gamma = 1$, and each bin is weighted by the expected black hole number. We have also explicitly indicated the semimajor axis range spanned by each grid to illustrate the hierarchy of the contributions: at low energy, we find most of the boosted DM flux is driven by the sizable population of black holes in long-period orbits. At high energy, by contrast, the flux becomes increasingly dominated by a considerably smaller number of black holes on short-period orbits, as these move fast enough to produce hard ejections. In fact, only about $30 - 40$ black holes dominate the flux at the highest ejection energy we have been able to numerically resolve. We emphasize that this hierarchy of contributions to the ejection spectrum is partially driven by the scaling of the DM and black hole density profiles, rather than being solely an intrinsic property of the boosting mechanism.

\begin{figure*}[t!]
    \centering
    \includegraphics[width=0.9\textwidth]{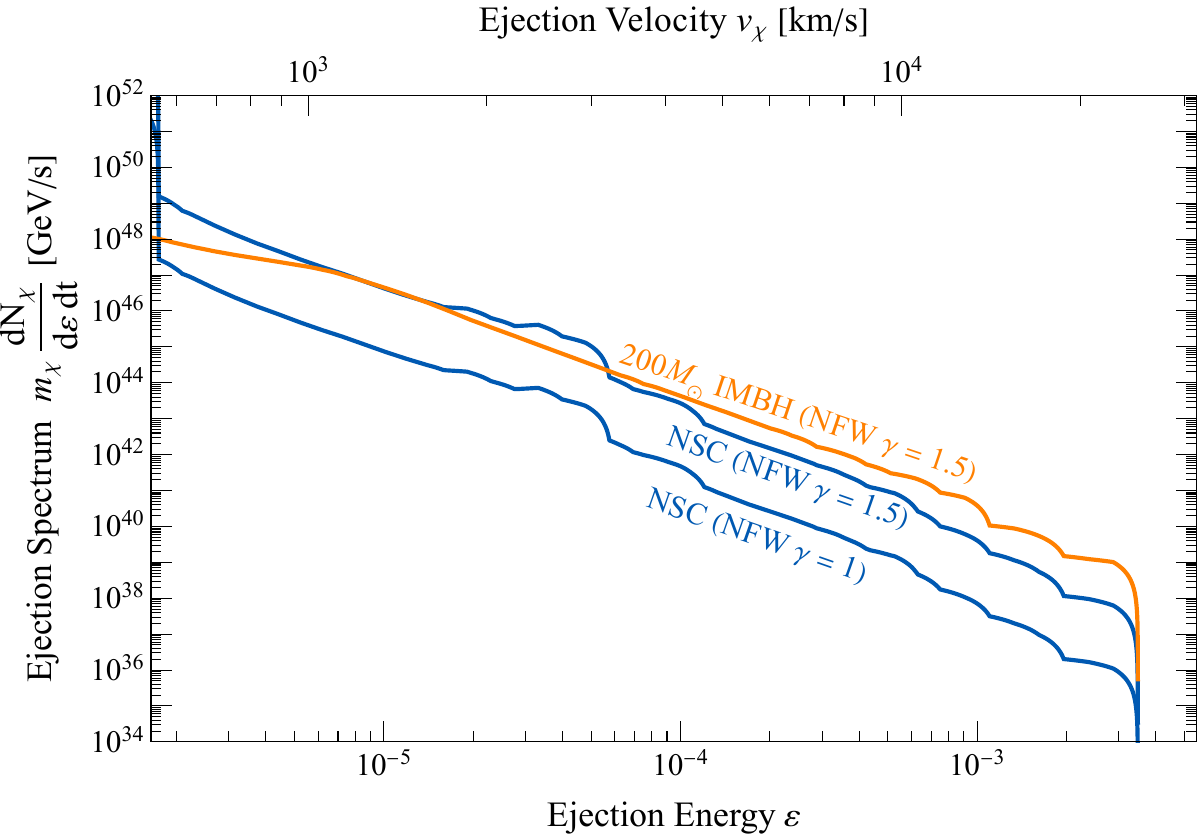}
    \caption{The full numerical ejection spectrum of the nuclear star cluster (\textit{blue}), obtained by summing over the contributions from every orbital element bin shown in Fig.~\ref{fig:sim_grid_BW}. The two lines bracket the variation associated with the NFW slope at the boundary of this region. We additionally show the contribution from a single intermediate-mass black hole on the shortest-period bin, assuming the steepest NFW profile (\textit{orange}).}
    \label{fig:Full_Spectrum}
\end{figure*}

Figure~\ref{fig:Full_Spectrum} shows the full ejection spectrum from the nuclear star cluster, obtained from Eq.~\eqref{eq:integrated_spec} by stacking the bins from every grid in Fig.~\ref{fig:sim_grid_BW}. We have bracketed the uncertainty associated with the halo profile by considering the full range of NFW slopes. As an additional benchmark, we have included the spectrum from a \textit{single} intermediate-mass black hole of $200 \, M_\odot$ in the shortest-period bin of the simulation grid, assuming the steepest NFW profile at the nuclear star cluster boundary. This curve is obtained from Eq.~\eqref{eq:integrated_spec} as well, by retaining only the corresponding term, fixing $N_{a,e} = 1$, and rescaling the spectrum quadratically with mass. Compared to the stellar-mass black hole population, the ejection spectrum in this case is mildly suppressed at low energies and modestly enhanced at high energies. The enhancement reflects both the approximate quadratic scaling of the ejection rate with companion mass, and our placement of the intermediate-mass black hole in the shortest-period bin, which dominates the highest-energy ejecta. As a result, we find that a single heavier object can produce a boosted DM flux comparable to the full population in the high-energy tail, despite the much lower occupation number.

By considering an intermediate-mass black hole, in addition to the full stellar-mass population, we seek to illustrate the degeneracy of our results with the assumed mass spectrum and black hole population number. It is worth noting that the existence of intermediate-mass black holes in the Milky Way's nuclear star cluster remains an open question. Current observational constraints are derived from precision astrometry of stars orbiting Sgr.~A$^\star$, and cannot rule out the presence of a dark massive perturber of up to $\sim 2000 \, M_\odot$ at the orbital distances we consider here \cite{2023A&A...672A..63G}. Our benchmark therefore lies comfortably below these existing limits. Future gravitational-wave measurements with LISA, under certain conditions, may further probe the population of intermediate-mass black holes in galactic nuclei, although the direct detectability of an isolated object with the mass and orbital parameters adopted here remains uncertain \cite{Strokov:2023kmo}.

To derive the velocity distribution, each orbital element bin acts as an independent and point-like source of boosted DM. The point-like approximation is justified by the enormous distance to the nuclear star cluster $D_{\rm NSC} \simeq 8.2 \ \rm kpc$ relative to the source's intrinsic size. For each individual bin, the associated velocity distribution is given by
\begin{equation}
    f_{a,e}(\mathbf{v_\chi}) = \mathcal{N}^{-1}_{\rm a,e} \, \frac{dp}{dv_\chi}\bigg|_{a,e}  \delta^{(2)}\!\!\left(\Omega_{v_\chi} - \Omega_0\right) \ , 
    \label{eq:vel_dist_NSC}
\end{equation}
\begin{equation}
    \frac{dp}{dv_\chi}\bigg|_{a,e} = v_\chi \, \, \frac{dp}{d\varepsilon} \bigg|_{a,e} \! \! (\varepsilon = v^2_\chi/2)~.
\end{equation}
The delta function enforces the fact that the observed velocity is collapsed onto a measure-zero set, where $\Omega_{v_\chi}$ denotes the direction in velocity space and $\Omega_0$ is the specific direction connecting the source and the observer.
The corresponding normalizing factor is 
\begin{equation}
    \mathcal{N}_{a,e} = \int d^3\mathbf{v_\chi} \, \frac{dp}{dv_\chi}\bigg|_{a,e} \, \delta^{(2)}\!\!\left(\Omega_{v_\chi} - \Omega_0\right)~,
\end{equation}
where $d^3\mathbf{v_\chi} = v_\chi^2 \, dv_\chi \, d\Omega_{v_\chi}$. We additionally require the number density to compute direct detection rates, which we obtain as 
 \begin{equation}
    \rho^{(a,e)}_\chi = m_\chi \,N_{a,e} \int_0^{\infty} \frac{1}{v_\chi} \frac{d^3N_\chi}{dA \, dv_\chi \, dt}\bigg|_{a,e} \, dv_\chi = \frac{m_\chi \, N_{a,e}}{4 \pi D_{\rm NSC}^2} \int_0^\infty \mathcal{F}^{(a,e)}_{\rm sim} \, \frac{dp}{d\varepsilon}\bigg{|}_{a,e} \frac{d\varepsilon}{\sqrt{2 \varepsilon}}~.
\end{equation}
On the rightmost side above, we have converted the integrand to the same form as the terms in Eq.~\eqref{eq:integrated_spec}, by using $dA = D_{\rm NSC}^2 \, d\Omega$ for the differential area, and $d\varepsilon = v_\chi \, dv_\chi$ for the differential energy. The full velocity distribution is then constructed by stacking every $f_{a,e}(\mathbf{v}_\chi)$, weighted by its corresponding mass density $\rho^{(a,e)}_\chi$. This produces a velocity distribution that is normalized to the total boosted DM density sourced by the nuclear star cluster at Earth's location. 

\begin{figure*}[t!]
    \centering
    \includegraphics[width=0.9\textwidth]{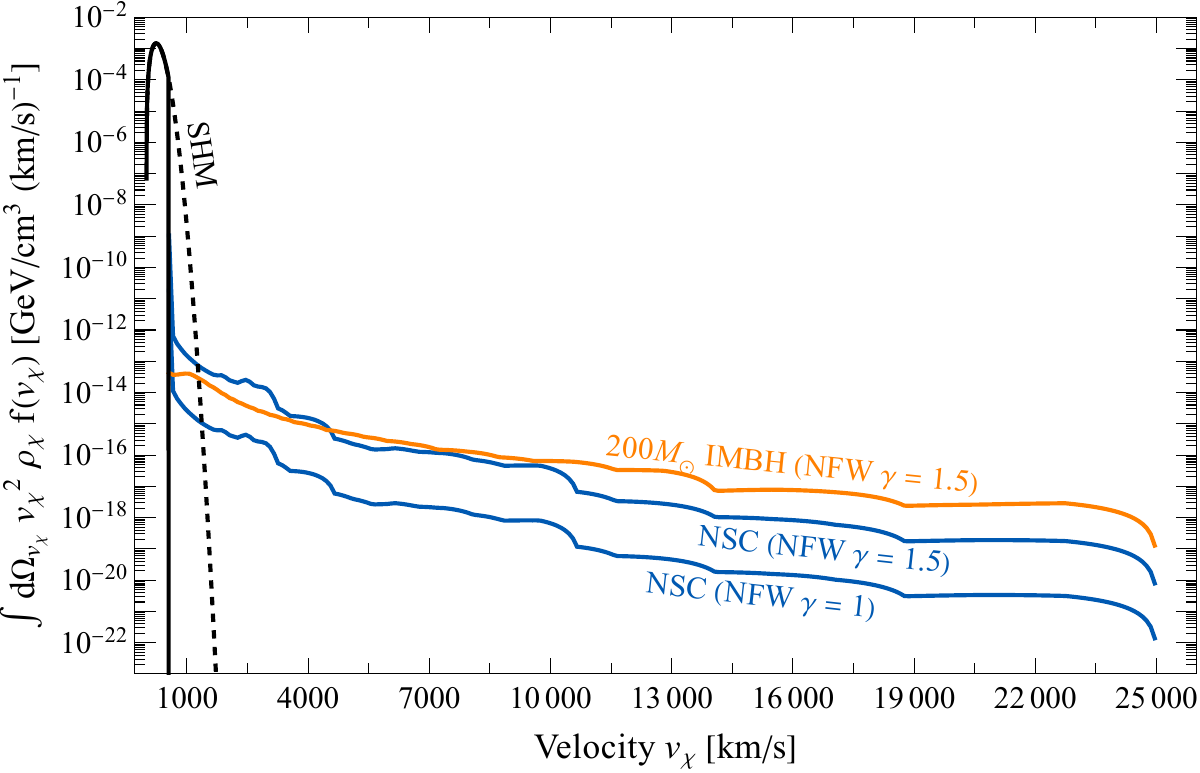}
    \caption{The velocity distribution for the ejected DM. We show the contribution from stellar-mass black holes, assuming an NFW profile with varying inner slope at the boundary of the nuclear star cluster (\textit{blue}), as well as the contribution from a single intermediate-mass black hole populating the shortest-period orbit we have simulated (\textit{orange}). In the latter case, we have assumed the steepest NFW profile. For comparison, we have also plotted the Standard Halo Model distribution with (\textit{black}) and without (\textit{dashed black}) the conventional escape velocity cutoff.}
    \label{fig:Vel_dist}
\end{figure*}

Figure~\ref{fig:Vel_dist} shows the resulting velocity distribution, for the stellar-mass black hole population as well as the intermediate-mass black hole benchmark. For comparison, we have included the Standard Halo Model (SHM), which is given by
\begin{equation}
    f_{\rm SHM}(\mathbf{v}_\chi) = \mathcal{N}^{-1}_{\rm SHM} \ \rho_\chi \, \exp\left(-\frac{v_\chi^2}{\sigma_\chi^2}\right) \, \Theta(v_{\rm Gal} - v_\chi)~,
    \label{eq:vel_dist_SHM}
\end{equation}
\begin{equation}
    \mathcal{N}_{\rm SHM} = \, (\pi \sigma^2_\chi)^{3/2} \left[{\rm erf}\left(\frac{v_{\rm Gal}}{\sigma_\chi}\right) - \frac{2}{\sqrt{\pi}} \, \frac{v_{\rm Gal}}{\sigma_\chi} \, \exp\left(-\frac{v_{\rm Gal}}{\sigma_\chi}\right)^2\right]~,
    \label{eq:vel_dist_SHM_norm}
\end{equation}
where we have fixed $\sigma_\chi = 240 \ \rm km/s$ and the galactic escape velocity $v_{\rm Gal} \simeq 546 \ \rm km/s$ \cite{Folsom:2025lly}. We have also assumed $\rho_\chi = 0.42 \ \rm GeV/cm^3$ for the local DM density. The dashed line illustrates the exponential decay of the SHM if the galactic escape cutoff $v_{\rm Gal}$ is removed. Relative to the SHM, our estimated ejection spectrum reaches up to $\sim 100$ standard deviations higher in velocity, albeit with a much lower flux. As we show below, while this results in a sizable cross-section sensitivity gap between the SHM and the nuclear star cluster's contribution, this gravitationally-boosted DM component nevertheless can turn large-volume detectors into a competitive probe of light DM. 

To conclude this section, we point out a few implicit but largely inconsequential assumptions we have made in the calculation above. 
First, we have ignored the relative velocity between the nuclear star cluster and the DM halo. In fact, the velocity of Sgr.~A$^\star$ relative to the dynamical center of the Milky Way has been recently observed to be $\lesssim 1 \ \rm km/s$ \cite{2020ApJ...892...39R}. Assuming for simplicity that the Milky Way's DM halo is aligned and at rest with the galaxy's dynamical center, this implies an almost negligible DM bulk motion as seen from nuclear star cluster's rest frame. 
Furthermore, as argued in Ref.~\cite{Acevedo:2026xol}, including a bulk velocity in the DM distribution has a mild impact on the ejection spectra, even when it is comparable to the center-of-mass velocity of the simulated system. 
Second, we have also ignored any gravitational potential gradient between the source and the observation point. In terms of escape velocity, the difference between the boundary of the nuclear star cluster and the Solar System position is estimated to be of order $\sim 200 \ \rm km/s$ among the various Milky Way mass models \cite{2013A&A...549A.137I}. This potentially only slows down the lowest-energy ejecta into the SHM-dominated range, which do not dominate our signal in any case. 
Finally, this calculation assumes the instantaneous ejection spectrum at the source matches the spectrum at detection. This is valid so long as the travel time of the ejecta along the baseline is small compared to the evolution timescale of the orbiting black hole. We provide an overview of various timescales associated with the orbital evolution in App.~\ref{app:timescales}, but we note here that they have minimal impact on the numerical spectrum as computed. 

\section{Direct Detection Prospects}
\label{sec:DD_prospects}
We now proceed to compute the extended sensitivity of dark matter direct detection experiments resulting from the boosted DM flux sourced by the nuclear star cluster. 

We must first convert the velocity distribution derived previously into the detector's rest frame. This is done by boosting $v_\chi \rightarrow |\mathbf{v}_\chi+\mathbf{V}_{\rm det}|$, where $\mathbf{V}_{\rm det}$ is the velocity of the detector relative to the halo. This is typically parameterized as $\mathbf{V}_{\rm det} = \mathbf{V}_{\rm \odot} + \mathbf{V}_{\rm \oplus}(t)$, where there is a fixed contribution from the Sun's motion around the Milky Way, and a periodic contribution from the Earth's motion around the Sun. We neglect the latter, as the amplitude of $|\mathbf{V}_{\oplus}(t)| \simeq 30 \ \rm km/s$ is small compared to $|\mathbf{V}_{\odot}| \simeq 232 \ \rm km/s$.

For a specified differential cross-section $d\sigma/dE_R$, the differential recoil rate is 
\begin{equation}
    \frac{dR}{dE_R} = N_T \int_{v_\chi > v^{\rm (min)}_\chi} \eta(E_R) \,  \frac{d\sigma}{dE_R} \, v_\chi \, f(\mathbf{v}_\chi) \, d^3 \mathbf{v}_\chi~,
    \label{eq:diff_recoil_rate}
\end{equation}
where $N_T$ is the total number of targets in the detector, and $f(\mathbf{v}_\chi)$ is the velocity distribution from hereon converted to the detector frame and normalized to DM number density. The function $\eta(E_R)$ is the detector-specific efficiency for observing a nuclear recoil of energy $E_R$. Eq.~\eqref{eq:diff_recoil_rate} is formally integrated for all velocities starting from some minimum value $v_\chi^{\rm min}$ kinematically compatible with recoil energy $E_R$, which is specified further below for both the elastic and inelastic cases.

We obtain the expected event number associated with DM interactions by integrating Eq.~\eqref{eq:diff_recoil_rate} over all recoil energies,
\begin{equation}
    N_\chi = \mathcal{T} \, \int_0^{\infty} \frac{dR}{dE_R} \, dE_R~,
    \label{eq:event_number}
\end{equation}
where $\mathcal{T}$ is the exposure. While we wrote the above as formally integrated over all recoil energies, in practice there is an upper bound on $E_R$ set by either the detector's efficiency $\eta(E_R)$ or some maximum velocity for the transiting DM. 

For the remainder of the analysis, we consider the four most sensitive large-volume detectors currently in operation: LZ, PandaX-4T, XENONnT, and DarkSide-50. Each of these experiments has performed light DM searches, relying only on the ionization component of the scattering signal \cite{LZ:2025igz,PandaX:2025rrz,DarkSide-50:2022qzh,XENON:2026qow}. Relaxing the requirement of a paired scintillation-ionization event has allowed these collaborations to reach sub-keV detection thresholds and, subsequently, probe DM-nucleon scattering for masses in the $1-3$ GeV scale, albeit with higher instrumental background and detector response uncertainty. We focus on these searches for elastic spin-independent and spin-dependent interactions. For the inelastic high-mass scenario, however, we instead restrict ourselves to the standard WIMP recoil energy window, which provides the dominant sensitivity in that regime.

Specifically, we are interested in comparing the reported limit under the SHM to the enhanced sensitivity obtained from Sgr.~A$^\star$'s boosted DM contribution. To this end, we separately compute the differential recoil rate using either $f_{\rm SHM}$ or $f_{\rm NSC}$ as the input distribution. We then solve for the cross-section sensitivity in Eq.~\eqref{eq:event_number}, given a number of events $N_\chi$ determined by the statistical significance. 
For LZ, we utilize their reported upper bound on the number of DM events during their run. 
For PandaX-4T, instead we derive the $1\sigma$ confidence level limit based on the procedure outlined in Ref.~\cite{Raj:2024guv}, which we use to fix $N_\chi$ above and solve for the cross-section assuming $N_{\rm obs} \simeq N_{\rm bkg} = 322$ as the number of observed/background events and $\sigma_{\rm bkg} \simeq 0.1$ as the background uncertainty. 
Similarly, for XENONnT, we use the same procedure assuming instead $N_{\rm obs} \simeq N_{\rm bkg} = 2554$ and $\sigma_{\rm bkg} \simeq 0.035$. 
For DarkSide-50, there is the additional caveat that their detection efficiency is reported as a function of photoelectrons. In this case, we convert nuclear recoil energy to electron yield using the ionization response function from Ref.~\cite{DarkSide:2021bnz}, and reformulate Eq.~\eqref{eq:event_number} as a discrete sum over the number of produced electrons per nuclear scattering. The sum is limited to $\geq 4$ electrons produced per scattering event, per the analysis of Ref.~\cite{DarkSide-50:2022qzh}. We then apply the same procedure as the two previous experiments with $N_{\rm obs} \simeq N_{\rm bkg} = 432,264$ and $\sigma_{\rm bkg} \simeq 0.1$. 
In each case, when assuming the SHM distribution, we reproduce the reported limits to within a factor $\sim 2-3$. We consider this level of discrepancy acceptable for the purposes of this work, as our goal is to compare relative sensitivities rather than deriving precise constraints.

\begin{figure}[t!]
\centering
\vspace{-0.9cm}
\hspace{-0.8cm}
\begin{subfigure}{0.51\textwidth}
    \includegraphics[width=\linewidth]{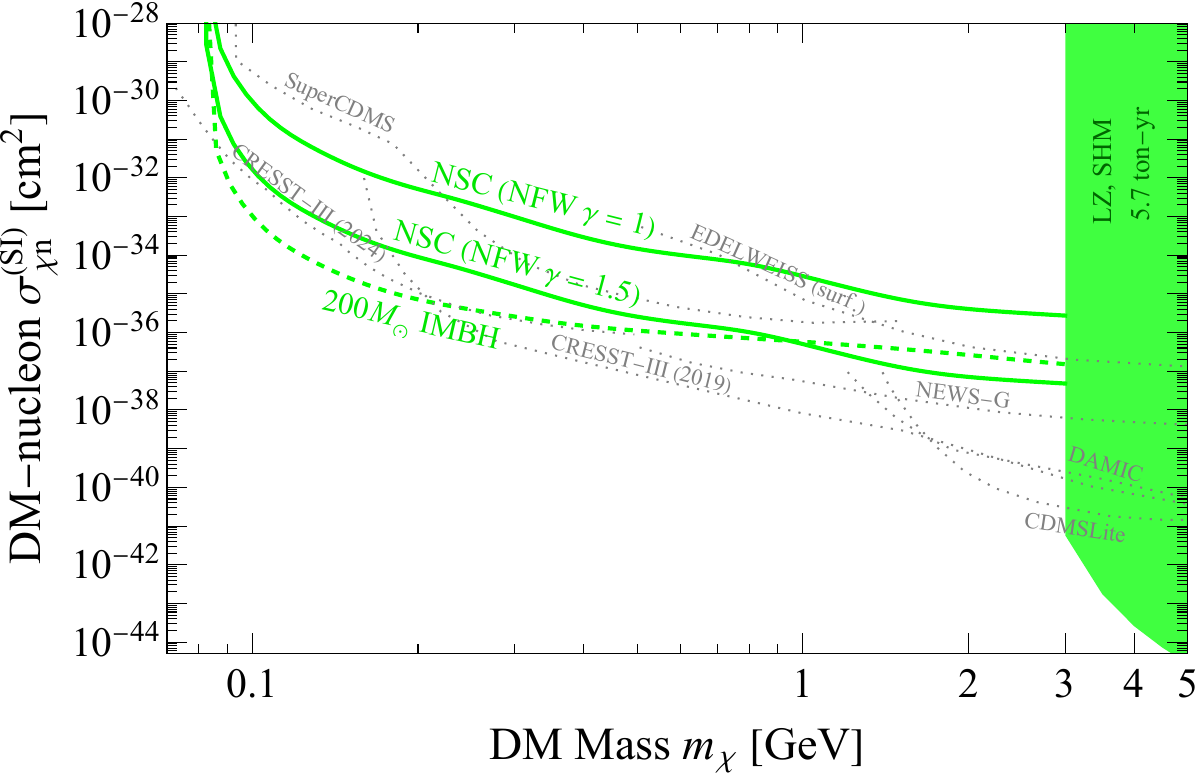}
\end{subfigure}
\hfill
\begin{subfigure}{0.51\textwidth}
    \includegraphics[width=\linewidth]{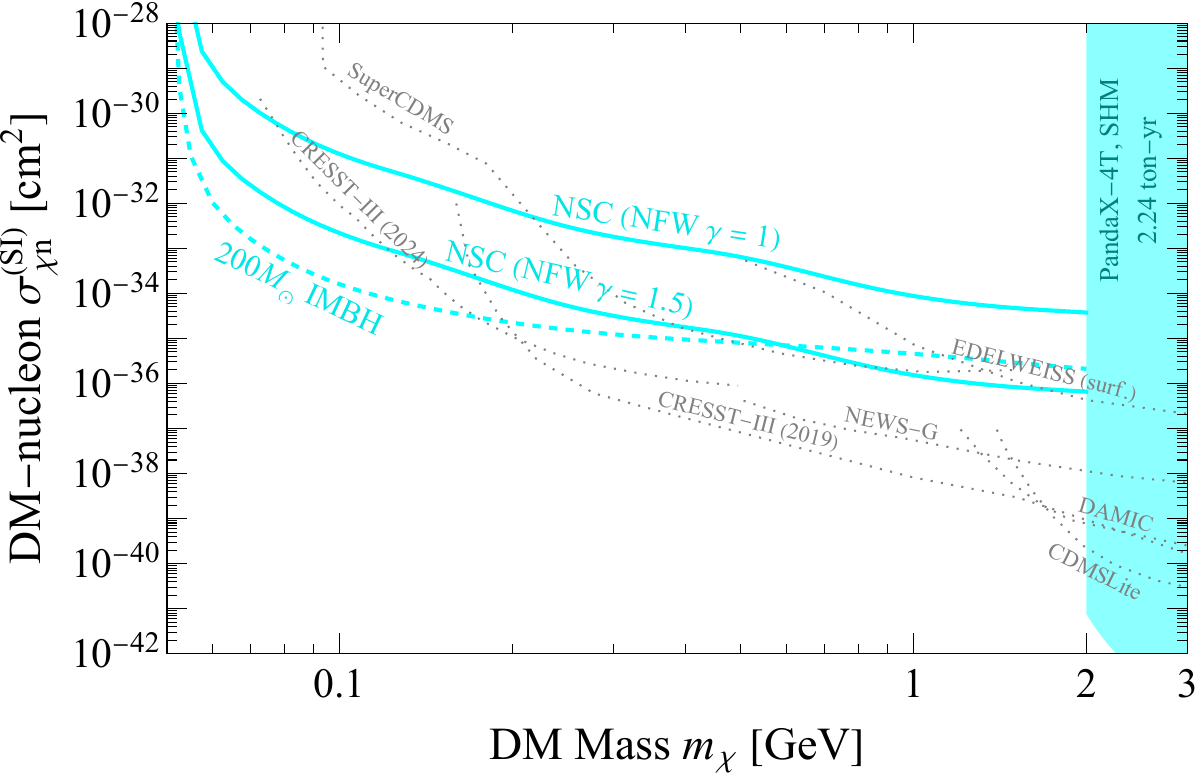}
\end{subfigure}
\vspace{0.1cm}
\hspace{-0.8cm}
\begin{subfigure}{0.51\textwidth}
    \includegraphics[width=\linewidth]{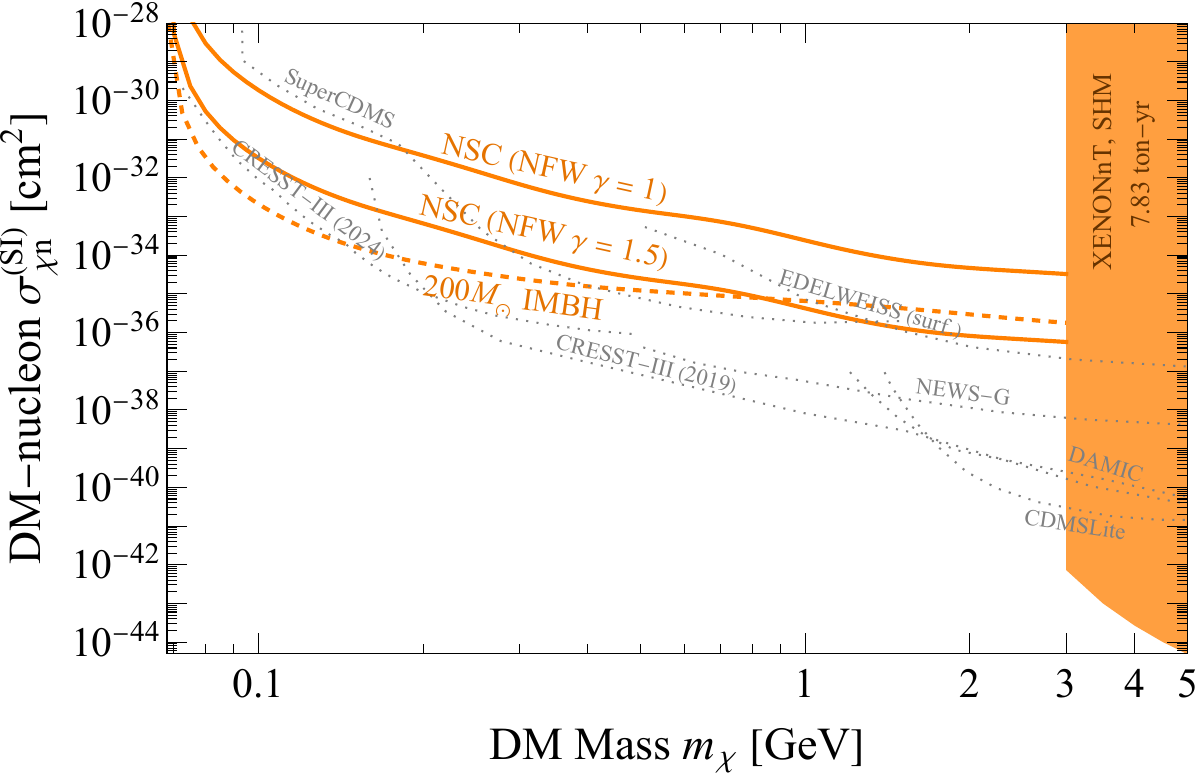}
\end{subfigure}
\hfill
\begin{subfigure}{0.51\textwidth}
    \includegraphics[width=\linewidth]{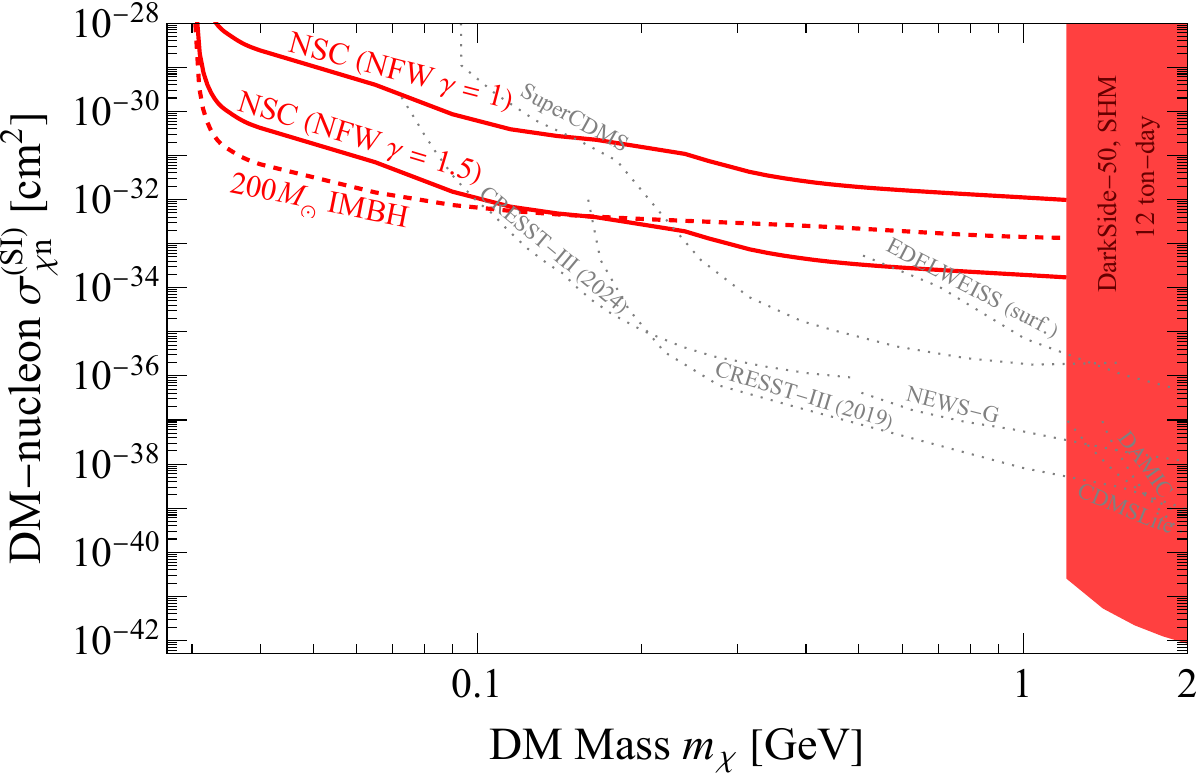}
\end{subfigure}
\caption{The extended mass sensitivity for spin-independent DM-nucleon scattering for LZ (\textit{top left/green}), PandaX-4T (\textit{top right/cyan}), XENONnT (\textit{bottom left/orange}) and DarkSide-50 (\textit{bottom right/red}). Solid lines denote the contribution from stellar-mass black holes, assuming an NFW profile with varying inner slope at the boundary of the nuclear star cluster. The dashed line shows the contribution from a single intermediate-mass black hole on the shortest-period orbit we have simulated, assuming the largest NFW slope shown.
The shaded region in each case shows the experiment's current limit under the Standard Halo Model. Additional constraints from NEWS-G \cite{NEWS-G:2017pxg}, CRESST \cite{CRESST:2019jnq,CRESST:2024cpr}, DAMIC \cite{DAMIC:2020cut}, CDMSLite \cite{SuperCDMS:2017nns}, and above-ground searches from EDELWEISS \cite{EDELWEISS:2019vjv} and SuperCDMS \cite{SuperCDMS:2020aus} under the SHM (\textit{gray dashed}) are shown for comparison.}
\label{fig:SI}
\end{figure}

We emphasize that, for simplicity, we have limited the scope of this analysis to DM-nucleus scattering. An analogous study of DM-electron scattering would be more involved as it would require detailed modeling of the electronic response function of the scattering target. Moreover, the resulting sensitivity extension possibly probes DM masses and cross-sections already constrained by various existing stellar cooling and cosmological constraints \cite{Hardy:2016kme,An:2013yfc,Viel:2013fqw,Sabti:2019mhn,Buen-Abad:2021mvc}.

\subsection{Spin-independent Scattering}
We start by considering elastic spin-independent interactions, for which we use the standard form for the differential cross-section,
\begin{equation}
    \frac{d\sigma_{\chi N}^{\rm (SI)}}{dE_R} = \frac{\sigma^{\rm (SI)}_{\chi n}}{\mu^2_{\chi n}} \frac{m_N}{2 v_\chi^2} A^2 \, F_{\rm SI}^2(q) ~.
    \label{eq:SI-cross-section}
\end{equation}
Above, $A = 131$ ($A = 40$) is the nuclear mass number for xenon (argon), $m_N$ is the corresponding nuclear mass, $\mu_{\chi n}$ is the DM-nucleon reduced mass, and $\sigma^{\rm (SI)}_{\chi n}$ is the spin-independent reference cross-section at the nucleon level. The Helm form factor $F_{\rm SI}^2(q)$ incorporates the loss of scattering coherence at large momentum transfers, and is parameterized by \cite{Duda:2006uk}
\begin{equation}
    F_{\rm SI}^2(q) = \frac{3j_1(q \, r_N)}{q \, r_N} \ \exp\left[-\frac{(q \, s_N)^2}{2}\right]~,
\end{equation}
where $j_1(x)$ is the spherical Bessel function of the first kind, $r_N \simeq (1.14 \ {\rm fm}) \, A^{1/3}$ is the nuclear radius, $s_N \simeq 0.9 \ \rm fm$, and $q = 2 m_N E_R$ is the momentum transfer magnitude. For elastic scattering, Eq.~\eqref{eq:diff_recoil_rate} is integrated from a minimum velocity compatible with recoil energy $E_R$ given by
\begin{equation}
    v_{\chi}^{\rm (min)}(E_R) = \sqrt{\frac{m_N E_R}{2 \mu^2_{\chi N}}}~,
    \label{eq:vmin_elastic}
\end{equation}
where $\mu_{\chi N}$ is the DM-nucleus reduced mass.
\begin{figure*}[t!]
    \centering
    \includegraphics[width=0.9\textwidth]{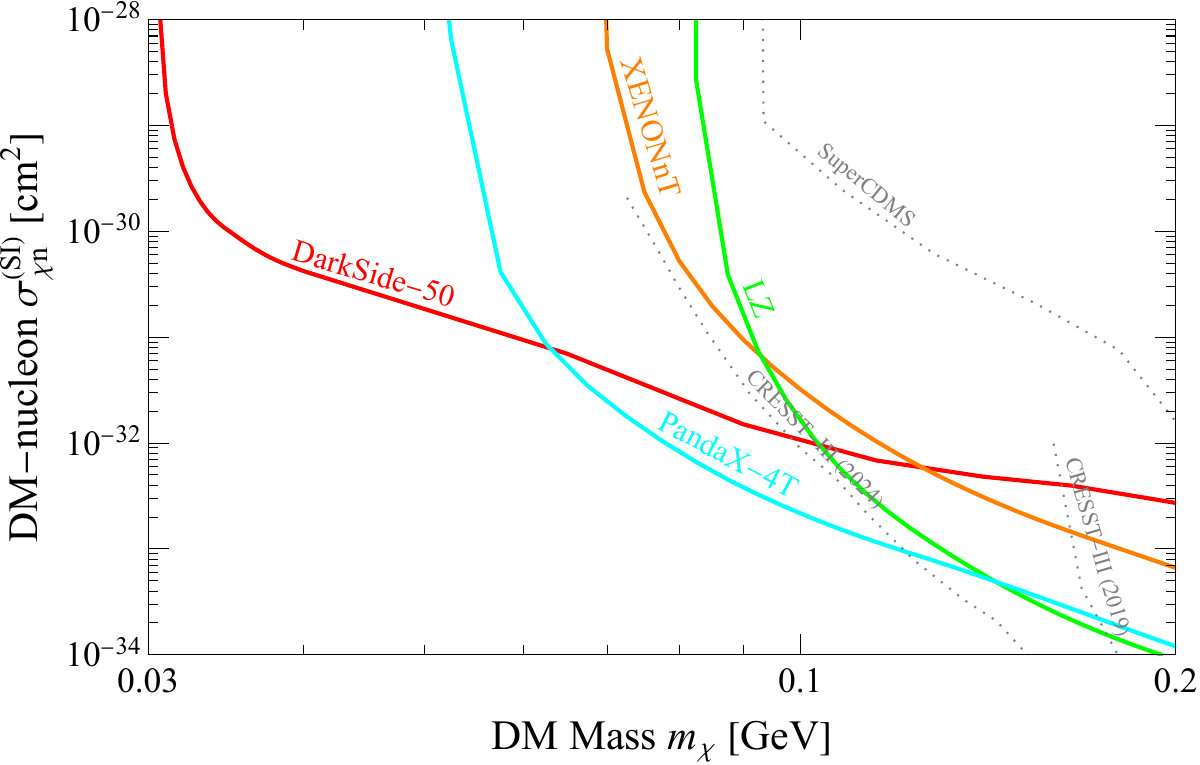}
    \caption{Combined sensitivity projections from Fig.~\ref{fig:SI} for all large-volume detectors, as indicated, focusing on the low mass regime. Each contour corresponds to the contribution from stellar-mass black holes in the nuclear star cluster, assuming an NFW profile with slope $\gamma = 1.5$ at the boundary of this region.}
    \label{fig:SI-combined}
\end{figure*}

Figure~\ref{fig:SI} shows the extended mass sensitivity of each experiment to spin-independent nuclear scattering. For comparison, we have plotted in each case the corresponding limit assuming the SHM with the standard galactic escape cutoff, as well as several additional constraints reported by various low-threshold and/or surface-operated experiments assuming also the SHM. The boosted flux produced by Sgr.~A$^\star$'s orbiting companions extends the mass sensitivity of xenon-based detectors down to $\sim 60 \ \rm MeV$, and $\sim 30 \ \rm MeV$ for argon-based DarkSide-50. The latter, in particular, can significantly outperform low-threshold detectors at this mass scale thanks to its use of a lighter nuclear target, independently of the assumed DM halo profile. Xenon-based detectors, on the other hand, can reach cross-sections competitive with CRESST and SuperCDMS, depending on the assumed halo profile. PandaX-4T is able to reach marginally lighter masses than CRESST, as this experiment achieved the lowest threshold energy in their search compared to LZ or XENONnT. Interestingly, even a single intermediate-mass black hole bound in close proximity to Sgr.~A$^\star$ is able to produce qualitatively similar results. In this scenario, the enhancement in the rate of high energy ejecta translates directly into an increased cross-section sensitivity at the lightest masses.   

We terminate the sensitivity lines at a cross-section $\sim 10^{-28} \ \rm cm^2$. Above this approximate value, energy loss of the DM particles as they cross the rock overburden can no longer be neglected as we have done here. This is also why we do not perform the same analysis for the low-threshold detectors we show: their smaller exposure necessarily implies the sensitivity extension will occur at much larger cross-sections, which will already lie close to or above the overburden threshold. Consequently, the gain in mass reach will be more limited in these cases. However, it is worth noting that this analysis could be applied to future upgrades of these experiments when they achieve a higher cross-section sensitivity. 

Figure~\ref{fig:SI-combined} shows a zoom-in of the DM mass range $(30 - 200) \ \rm MeV$ with all detectors in Fig.~\ref{fig:SI} combined, in order to allow direct comparison between the four experiments, as well as to further highlight the potential sensitivity gain. For simplicity, in this case we have only shown the contribution from stellar-mass black holes and an NFW profile with $\gamma = 1.5$ as the boundary condition. 

In this presentation, we have not included experimental limits derived under additional assumptions, such as the Migdal effect \cite{DarkSide:2022dhx,PandaX:2023xgl,CDEX:2019hzn,XENON:2019zpr} or bremsstrahlung \cite{XENON:2019zpr}, to facilitate a comparison under the most minimal framework. In practice, such additional assumptions could also be incorporated into our analysis, given its largely model-independent nature.

\subsection{Spin-dependent Scattering}
We next consider the case of spin-dependent DM-nucleus scattering, with a differential cross-section parameterized as
\begin{equation}
    \frac{d\sigma_{\chi N}^{\rm (SD)}}{dE_R} = \frac{\sigma^{\rm (SD)}_{\chi (p,n)}}{\mu^2_{\chi n}} \frac{m_N}{2 v_\chi^2} \frac{4 \pi }{3 (2 J + 1)}\, S_{\rm (p,n)}(q)~.
    \label{eq:SD-cross-section}
\end{equation}
Above, $J$ is the nuclear target's spin and, as before, $m_N$ is the nuclear mass, $\mu_{\chi n}$ is the DM-nucleon reduced mass, and $v_\chi$ is the DM velocity in the detector's frame. 
The function $S_{(p,n)}(q)$ is the corresponding momentum transfer–dependent nuclear structure function, which we obtain from Ref.~\cite{Klos:2013rwa}, and $\sigma^{\rm (SD)}_{\chi (p,n)}$ is the reference nucleon-level cross section at zero momentum transfer. 
The additional index $(p,n)$ indicates whether scattering is dominated by the proton or neutron spin. We assume this process is also elastic, so the same kinematic limits for computing the recoil rate apply as in the prior case, $cf.$ Eq.~\eqref{eq:vmin_elastic}. 

We focus on PandaX-4T for this benchmark interaction. The corresponding analysis for LZ and XENONnT would produce qualitatively similar results, although their extended mass reach would be somewhat lower relative to PandaX-4T. We do not consider DarkSide-50, as argon is insensitive to spin-dependent scattering at the tree-level, owing to the absence of unpaired neutrons or protons \cite{Bozorgnia:2024kkf}. As with the spin-independent scenario above, other experiments with lower detection thresholds have exposures too small to access new parameter space without simultaneously requiring cross-sections too large for the overburden to be neglected.

\begin{figure*}[t!]
    \centering
    \includegraphics[width=0.9\textwidth]{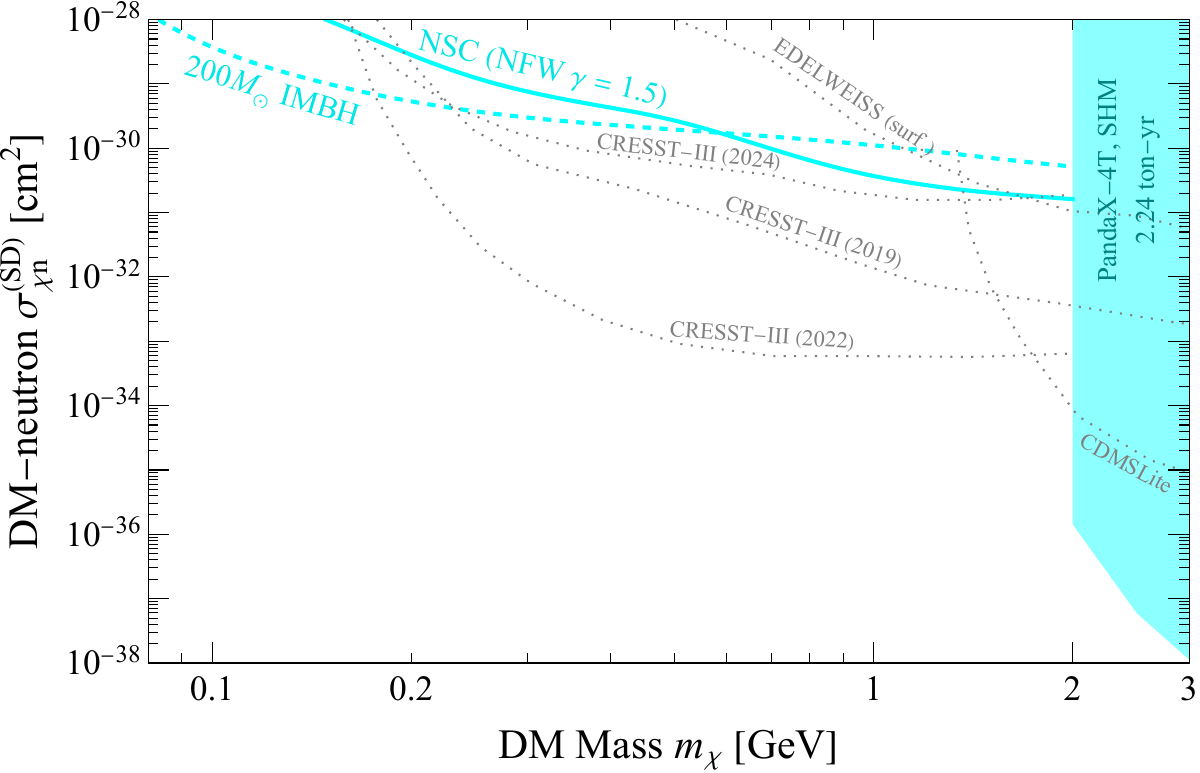}
    \caption{The extended mass sensitivity for spin-dependent neutron-only scattering for PandaX-4T (\textit{cyan}). The solid line denotes the contribution from stellar-mass black holes, while the dashed line shows the contribution from a single intermediate-mass black hole on the shortest-period orbit simulated. Both are computed assuming the steepest NFW profile considered at the boundary of the nuclear star cluster. The shaded region shows the approximate current limit under the Standard Halo Model. Additional limits assuming the SHM from CRESST \cite{CRESST:2019jnq,CRESST:2022dtl,CRESST:2024cpr}, CDMSLite \cite{SuperCDMS:2017nns}, and an above-ground search performed by EDELWEISS \cite{EDELWEISS:2019vjv} are shown for comparison.}
    \label{fig:PandaX_SD_Neutron}
\end{figure*}

Figure~\ref{fig:PandaX_SD_Neutron} shows the extended sensitivity of PandaX-4T to neutron-only spin-dependent scattering. For the steeper NFW profile, the boosted DM flux from the stellar-mass black hole population allows this detector to reach masses as low as $\sim 150 \ \rm MeV$ and cross-sections comparable to various CRESST runs towards this value. Below this mass scale, the required cross-section becomes sufficiently large that overburden effects would become important. As with the prior case, a single intermediate-mass black hole on the smallest simulated orbit further improves sensitivity at low masses. Notably, in this scenario, the projected sensitivity surpasses current limits at the low-mass end, potentially rendering this detector a complementary probe of this interaction benchmark should such an intermediate-mass black hole be observed in the near future. For proton-only interactions, we only find limited mass sensitivity improvement assuming the upper end of our range of halo profile slopes, and so we do not show this case for simplicity. 

\subsection{Inelastic Scattering}
Finally, we analyze the inelastic process $\chi_{1} + N \rightarrow \chi_{2} + N$ where the initial and final DM states differ in mass by a splitting $\delta = m_{\chi_2} - m_{\chi_1}$. We treat this gap as an additional, independent parameter that is ultimately determined by the underlying DM particle model. Inelastic DM scenarios have been extensively explored in the literature as a possible explanation for the persistent null results from direct detection searches \cite{Tucker-Smith:2001myb,Chang:2008gd,Finkbeiner:2007kk,Pospelov:2007xh,Batell:2009vb,Ghorbani:2014gka,Zhang:2016dck,Alvarez:2019nwt,Hooper:2025fda}. As we show below, the boosted DM flux from the nuclear star cluster can substantially extend the sensitivity of large-volume detectors to the mass splitting between initial and final states.

Because the boost is gravitational in origin, we are free to consider nearly arbitrary DM masses without incurring any penalty in energy gain, a feature that is unique to this mechanism. While we could consider light DM similarly to the previous sections, to highlight this feature we focus on TeV-scale inelastic DM, a regime that is significantly challenging to probe by direct detection experiments. For simplicity, we exclusively consider the sensitivity of LZ based on their most recent WIMP DM search \cite{LZ:2024zvo}. Analogous searches by PandaX-4T \cite{PandaX:2024qfu} and XENONnT \cite{XENON:2025vwd} would achieve a comparable reach in mass splitting, though with a somewhat weaker cross-section sensitivity relative to LZ. DarkSide-50, on the other hand, would not be competitive with these experiments in this regime. This is because the maximum mass difference that can be excited scales with target mass when $m_\chi \gg m_N$. LZ, PandaX-4T, and XENONnT, owing to their use of xenon as scattering target, therefore possess an intrinsically higher splitting reach than argon-based DarkSide-50. 

We specifically consider endothermic scattering, where the final DM state is heavier than the initial one. This imposes a higher velocity threshold for scattering with recoil energy $E_R$ compared to an elastic collision. In the detector frame, assuming non-relativistic kinematics and $m_N \ll m_\chi$, this is given by \cite{Bramante:2016rdh}
\begin{equation}
    v^{\rm (min)}_{\chi}(E_R) \simeq \frac{1}{\sqrt{2 m_N E_R}} \left(\frac{E_R \, m_N}{\mu_{\chi N}}+\delta\right)~.
    \label{eq:vmin_inelastic}
\end{equation}
Eq.~\eqref{eq:vmin_inelastic} implies that, for a sufficiently large splitting $\delta$, this process is prohibited by kinematics, given the finite velocity range expected for halo DM. In such a case, the only detection alternatives are loop-level elastic scattering, where the additional loop allows the final DM state to match the initial one, or the reverse process, $i.e.$ exothermic scattering, which does not impose a minimum velocity. The former has a cross-section generically suppressed by the loop structure, rendering this channel inaccessible in most cases. The latter depends on the relative relic abundance of the states $\chi_1$ and $\chi_2$, and therefore on the specific particle model for DM and its cosmological history. However, there exist well-motivated models in which the abundance of $\chi_2$ is mostly depleted, see $e.g.$ Ref.~\cite{Batell:2009vb}, making endothermic scattering the only viable detection channel. We therefore focus on this class of models, for which gravitationally-boosted DM can significantly enhance detection prospects. In what follows, we also assume the interaction to be spin-independent and maintain the same cross-section parameterization as in Eq.~\eqref{eq:SI-cross-section}.

\begin{figure*}[t!]
    \centering
    \includegraphics[width=0.9\textwidth]{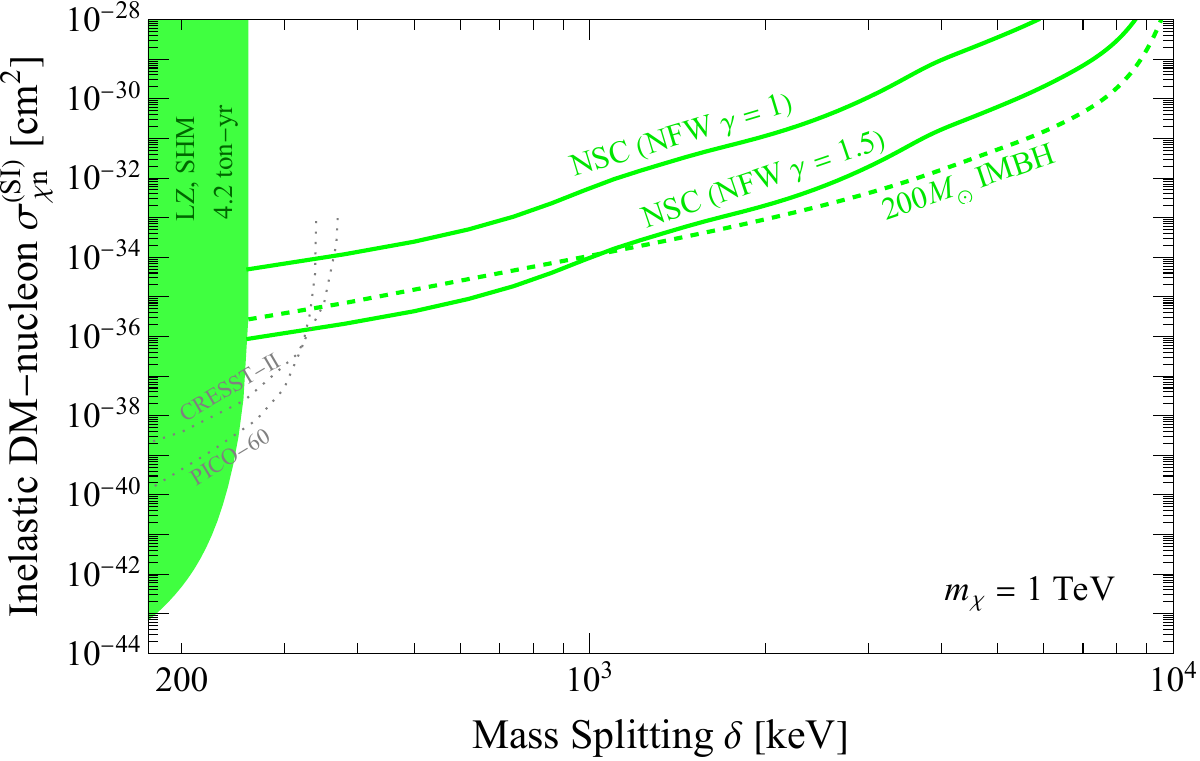}
    \caption{The extended mass splitting sensitivity for LZ (\textit{green}) in the inelastic scattering scenario. Solid lines denote the contribution from stellar-mass black holes, assuming an NFW profile with varying inner slope at the boundary of the nuclear star cluster. The dashed line shows the contribution from a single intermediate-mass black hole on the shortest-period orbit we have simulated, assuming the largest NFW slope shown. The shaded region denotes the experiment's sensitivity under the Standard Halo Model. Additional limits, also under the SHM, from PICO-60 and CRESST-II (\textit{gray}) are displayed for comparison \cite{Bramante:2016rdh}.}
    \label{fig:Inel_SI}
\end{figure*}

Figure~\ref{fig:Inel_SI} shows LZ's extended inelastic DM sensitivity for $m_\chi = 1 \ \rm TeV$, as a function of mass splitting. For comparison, we have also plotted limits from CRESST-II \cite{CRESST:2015txj} and PICO-60 \cite{PICO:2015pux} searches derived assuming the SHM only \cite{Bramante:2016rdh}. These two experiments are particularly powerful probes of inelastic DM thanks to their extended recoil energy search window in the case of PICO, and the use of tungsten as the scattering target in the case of CRESST. 
Depending on the assumed DM profile, the inclusion of the boosted DM flux from Sgr.~A$^\star$ allows LZ to probe up to $\sim 9 \ \rm MeV$ mass splittings before attenuation from the overburden becomes important.
The contribution from a single intermediate-mass black hole in close proximity to Sgr.~A$^\star$ can produce a modest improvement in sensitivity at large mass splittings. This mirrors the enhancement observed at low DM masses: both regions of parameter space probe the high-energy tail of the boosted DM spectrum, which is increased by an intermediate-mass black hole relative to the more numerous population of lighter black holes. 

It is worth noting that these sensitivity lines are largely independent of the underlying DM particle model up to cross-sections of $\sim 10^{-31} \, \rm cm^2$ \cite{Digman:2019wdm}. This transition occurs at around $\delta \simeq 2 \ \rm MeV$ ($\delta \simeq 3.5 \ \rm MeV$) for $\gamma = 1$ ($\gamma = 1.5$) NFW slopes when considering the full stellar-mass black hole population, and $\delta \simeq 5.5 \, \rm MeV$ for the intermediate-mass black hole scenario. 
Above this value, the cross-section scaling with nuclear mass number is no longer generic, although specific composite DM models that preserve it can be constructed \cite{Acevedo:2024lyr}. 
In any case, the boosted DM from the nuclear star cluster allows LZ to probe parameter space that is virtually untested by other direct detection experiments at this DM mass scale. In particular, the boosted DM flux from the nuclear star cluster renders detectors like LZ a considerably more robust probe than various existing or putative astrophysical searches in this parameter space \cite{McCullough:2010ai,Baryakhtar:2017dbj,Bell:2018pkk,Acevedo:2019agu,Acevedo:2025rqu,Acevedo:2024ttq,Alvarez:2023fjj,Gustafson:2025dff}.
We also remark that the splitting reach is partly limited by the experiment's maximum recoil energy acceptance of $\sim 70 \ \rm keV$, rather than the saturation of the kinematic threshold. Given the sizable ejection energies we observe, a wider recoil energy window could further improve both the cross-section and splitting reach. 

Lastly, considering the large flux and ejection energies we observe, we argue that large-volume neutrino detectors may provide another interesting venue for testing inelastic DM accelerated by Sgr.~A$^\star$'s orbiting companions. While their typical detection threshold is substantially larger compared to DM detectors, of order $\sim 10 - 100 \ \rm MeV$ instead of $\sim 1 - 10 \ \rm keV$, such a large DM mass combined with the boosts obtained allows for energy depositions onto nuclear targets of a similar level. In fact, the maximum nuclear recoil energy in a collision is given by (see $e.g.$ Ref.~\cite{Shu:2010ta}),
\begin{equation}
    E_R^{\rm max} = \frac{1}{2} \, m_\chi \, v_\chi^2 \left(1 - \frac{\mu^2_{\chi N}}{m^2_N} \left(1-\frac{m_N}{m_\chi} \sqrt{1-\frac{\delta}{\delta_{\rm max}}}\right)^2\right) - \delta~,
\end{equation}
\begin{equation}
   \delta_{\rm max} = \frac{1}{2} \, \mu_{\chi N} \, v_\chi^2 ~.
    \label{eq:deltamchi_max}
\end{equation}
Assuming an organic scintillator detector, we fix $m_N = 12 \ \rm GeV$, approximately the mass of a carbon nucleus. For the maximum observed ejection velocity of $\sim 25,\!000 \ \rm km/s$, this translates into $E^{\rm max}_R \gtrsim 150 \, \rm MeV$ when $\delta \lesssim \delta_{\rm max}/2$ for TeV DM. Therefore, the recoil energy range is compatible with the characteristic thresholds of this class of detectors. In contrast with the above, neutrino experiments would observe the high energy end of the recoil spectrum in a much wider energy window. Moreover, when accounting for the considerably larger volume (and therefore, exposure) relative to their DM counterparts, it is possible that the resulting inelastic cross-section sensitivity is significantly higher than compared to DM direct detection experiments, as obtained here. This relative gain, however, would likely be restricted to mass splittings $\gtrsim 0.5 \ \rm MeV$, where DM experiments lose sensitivity to endothermic scattering by halo particles. 
For smaller splittings, down to the elastic limit, this parameter space can be probed with high sensitivity by DM detectors, as illustrated in Fig.~\ref{fig:Inel_SI}. Given the fundamentally different characteristics of neutrino detectors relative to the DM direct detection experiments, we have left a detailed analysis of their reach for upcoming work \cite{AcevedoRitz:inprep}. 

\section{Conclusions}
\label{sec:conclusions}
In the present work, building on \cite{Acevedo:2026xol}, we have analyzed for the first time the gravitational ejection of halo DM by the stellar-mass black holes expected to reside in close proximity to Sgr.~A$^\star$. We find the immediate $\sim 0.2 \ \rm pc$ surrounding the central supermassive black hole constitutes the strongest galactic source of gravitationally-boosted DM, owing to its high DM density, large population of black holes, and deep gravitational potential. This provides a huge enhancement to both the flux and maximum energy reach of the ejecta compared to any other system in the Milky Way. Within our computational capabilities, we observe ejection velocities of up to $\sim 25,\!000 \ \rm km/s$. Although higher velocities are kinematically allowed, the extremely small probability of such ejection events prevented us from resolving them within the allotted simulation time.

Utilizing a well-motivated cusp profile for the black hole population, supported by theoretical works, available observational data and N-body simulations, we have derived the resulting velocity distribution for the boosted DM sourced by this region. We have shown that this flux can extend the sensitivity of large-volume detectors LZ, PandaX-4T, XENONnT and DarkSide-50 to DM masses far beyond what is accessible to them under the conventional halo velocity distribution. Moreover, we find that the boosted DM from this region renders these detectors competitive with lower-threshold experiments CRESST and SuperCDMS, which currently lie at the frontier of DM-nucleus scattering sensitivity in the sub-GeV regime. DarkSide-50, in particular, can access untested parameter space for spin-independent elastic scattering for DM at the $\sim 30 \ \rm MeV$ mass scale, thanks to its use of argon as the target material. Interestingly, an observed excess in this kinematic regime could admit a boosted DM interpretation, although a detailed assessment of alternative explanations would be required. More broadly, our work illustrates how astrophysical structures at the smallest scales may play a relevant role for DM direct detection.

Additionally, we have analyzed inelastic scattering for DM at the TeV scale. The gravitational nature of the boost uniquely enables access to this regime, since the energy gain of the up-scattered DM is independent of its mass. For LZ, the large ejection velocities characteristic of the nuclear star cluster enable sensitivity to endothermic scattering with mass splittings up to $\sim 3 \ \rm{MeV}$ in a largely model-independent manner, and up to $\sim 9 \ \rm{MeV}$ for larger cross-sections whose precise scaling depends on the underlying DM–SM interaction. In either case, this corresponds to previously unexplored regions of inelastic scattering parameter space. The reach in mass splitting is partly limited by the recoil energy window of this search; an increased upper bound on the recoil energy could further improve this result. The sizable boosted DM flux also opens the possibility of probing this class of inelastic models with neutrino detectors, which we leave for upcoming work.

These sensitivity projections are conservative in that we have not utilized aggressive choices for the DM profile, such as spike solutions. In fact, we have only considered ejections of unbound particles transiting through the nuclear star cluster, which represent a tiny fraction of the total DM content in this region. Moreover, we have not analyzed stronger mass segregation scenarios, whereby black holes preferentially populate the innermost region of the nuclear star cluster, which would boost ejections at high velocities and thus improve sensitivity to this fast-moving DM component at the lowest masses. The largest uncertainty is driven by the assumed black hole number, as this quantity sensitively depends on the assumed metallicity for this region. To illustrate the potential gain in sensitivity while remaining agnostic about the exact population, we have also considered a single, intermediate-mass black hole orbiting near Sgr.~A$^\star$. Even in this scenario, we have found that the resulting sensitivity improvement can convert large-volume detectors into a powerful probe of sub-GeV DM, as well as heavy inelastic DM. 

Moving forward, it will be interesting to consider whether future gravitational wave detectors can independently confirm the presence of black holes orbiting close to Sgr.~A$^\star$, thereby enabling sensitivity estimates based on confirmed sources rather than inferred population models. Together with future upgrades, as well as larger exposures, for the detectors considered here, the boosted DM flux from Sgr.~A$^\star$ may enable unprecedented sensitivity to otherwise challenging particle DM parameters. 

\acknowledgments
This research was supported in part by funding from the Natural Sciences and Engineering Research Council of Canada, including through the Arthur B. McDonald Canadian Astroparticle Physics Research Institute. This research was also enabled in part by support provided by the BC Digital Research Infrastructure Group, Compute Ontario (computeontario.ca), Calcul Qu\'ebec (calculquebec.ca), and the Digital Research Alliance of Canada (alliancecan.ca). Simulations were executed on the Alliance's Fir, Nibi and Rorqual high-performance computing clusters. 

\appendix

\section{Simulation Flux Factor}
\label{app:sim_factor}
In this appendix, we provide further details behind the derivation of Eq.~\eqref{eq:Fsim_gen} in the main text. We start by assuming that the unbound DM particles are characterized by a phase space distribution of the form
\begin{equation}
    \frac{d^2N_\chi}{d\mathbf{r} \,d\mathbf{v}} = f(\varepsilon) \propto \exp\left(-\frac{\varepsilon}{\varepsilon_0}\right)~,
    \label{eq:DF_DM}
\end{equation}
where, in contrast with Sec.~\ref{sec:NSC_mod}, here we define the energy such that $\varepsilon = v_\chi^2/2 \, + \, \phi(\mathbf{r}) > 0$ corresponds to unbound states. This corresponds to an isotropic velocity distribution when the gravitational potential vanishes,
\begin{equation}
    \frac{d^2N_\chi}{d\mathbf{r} \, d\mathbf{v}} \bigg|_{\infty} = f(\varepsilon = v_\chi^2/2) = \mathcal{N} \rho_\chi \exp{\left(-\frac{v^2_\chi}{2\, \varepsilon_0}\right)}~,\, \,  (v_\chi \geq 0)~.
    \label{eq:DF_DM_2}
\end{equation}
It is easy to see that $\varepsilon_0 = \sigma_\chi^2$ in order to identify Eq.~\eqref{eq:DF_DM_2} with the standard Maxwell-Boltzmann distribution (without any escape velocity truncation). The normalization constant here is accordingly 
$\mathcal{N}^{-1} = (2\pi)^{3/2} \sigma^3_\chi$, such that integration over velocity yields the background density $\rho_\chi$. 

By Liouville's theorem, the distribution function preserves its functional form in terms of the integrals of motion. Therefore, at the simulation boundary, it may be evaluated in the same manner, now including Sgr.~A$^\star$'s gravitational potential, 
\begin{equation}
    \frac{d^2N_\chi}{d\mathbf{r} \, d\mathbf{v}} \bigg|_{\rm sim} = f\left(\varepsilon = v_\chi^2/2 \, + \, \phi_{\rm sim}\right) = \mathcal{N} \rho_\chi \exp{\left(-\frac{v^2_\chi + 2\phi_{\rm sim}}{2 \sigma_\chi^2} \right)}~,\, \,  \left(v_\chi \geq \sqrt{2 \phi_{\rm sim}}\right)~,
\end{equation}
where we have defined for brevity $\phi_{\rm sim} = \phi(r = R_{\rm sim})$.
Relative to the distribution at infinity, the distribution at the simulation boundary absorbs a factor $\exp(-\phi_{\rm sim}/\varepsilon_0)$ in the normalization, and its domain is now limited to velocities above Sgr.~A$^\star$'s escape velocity. It is then possible to directly integrate over the simulation boundary to obtain $\mathcal{F}_{\rm sim}$,
\begin{equation}
    \mathcal{F}_{\rm sim} =  \int_{\Sigma_{\rm sim}} \left[\int f\left(v_\chi^2/2 \, + \, \phi_{\rm sim}\right) \, \mathbf{v}_\chi \, d\mathbf{v}_\chi\right] \cdot \hat{\mathbf{n}} \, dA~,
\end{equation}
where $\Sigma_{\rm sim}$ is the spherical simulation boundary centered at the position of Sgr.~A$^\star$. We parameterize $\mathbf{v}_\chi = v_\chi (\sin\theta_v \cos\varphi_v, \sin\theta_v \sin\varphi_v,\cos\theta_v)$, where the z-direction is defined parallel to $\hat{\mathbf{n}}$ at each point in the surface. The first two components vanish upon integrating over $\varphi_v$. The remaining component yields 
\begin{align}
    \int f\left(v_\chi^2/2 \, + \, \phi_{\rm sim}\right) \, \mathbf{v}_\chi \, d\mathbf{v}_\chi = \, \frac{\rho_\chi \exp\left(- \phi_{\rm sim}/\sigma_\chi^2\right)}{(2 \pi)^{3/2} \, \sigma_\chi^3} & \left[\int_{\sqrt{2 |\phi_{\rm sim}|}}^\infty 4 \pi v_\chi^3 \exp{\left(-\frac{v^2_\chi}{2 \sigma_\chi^2}\right)} dv_\chi\right] \times \hat{\mathbf{z}} \nonumber \\
   = \frac{4\pi \rho_\chi}{(2 \pi)^{3/2} \sigma_\chi}  \left(2 |\phi_{\rm sim}| + 2 \sigma_\chi^2 \right) \times \hat{\mathbf{z}}~.
\end{align}
Since this is parallel to the direction defined by $\hat{\mathbf{n}}$, the remaining integration over the boundary becomes equal to half its surface area $2\pi R^2_{\rm sim}$, where the additional factor $1/2$ rejects the outgoing flux, resulting in
\begin{equation}
    \mathcal{F}_{\rm sim} = \sqrt{\frac{2}{\pi}} \, 4 \pi R_{\rm sim}^2 \, \rho_\chi \, \sigma_\chi\left({1 + \frac{|\phi_{\rm sim}|}{\sigma^2_\chi}}\right)~.
    \label{eq:Fsim_gen_appendix}
\end{equation}
In the limit $R_{\rm sim} \rightarrow \infty$, the term containing $\phi_{\rm sim}$ vanishes, and Eq.~\eqref{eq:Fsim_gen_appendix} asymptotes to the standard flux formula,
\begin{equation}
    \mathcal{F}_{\rm sim} \simeq \, 4\pi R_{\rm sim}^2 \, \rho_\chi \, \langle v_\chi\rangle~,
\end{equation}
where $\langle v_\chi\rangle = \sqrt{2/\pi} \, \sigma_\chi$ for an undistorted Maxwellian distribution. Eq.~\eqref{eq:Fsim_gen_appendix} may be arranged in the same form, making the gravitational focusing of both the density and velocity dispersion more manifest,
\begin{equation}
    \mathcal{F}_{\rm sim} \simeq \, \sqrt{\frac{2}{\pi}}\, 4\pi R_{\rm sim}^2 \, \rho^{\rm sim}_\chi \, \sigma^{\rm sim}_\chi~,
\end{equation}
where $\rho^{\rm sim}_\chi$ and $\sigma^{\rm sim}_\chi$ are now the enhanced local density and velocity dispersion
\begin{equation}
    \rho^{\rm sim}_\chi = \rho_\chi \,\sqrt{1+\frac{|\phi_{\rm sim}|}{\sigma^2_\chi}}~,
\end{equation}
\begin{equation}
    \sigma^{\rm sim}_\chi = \sigma_\chi \,\sqrt{1+\frac{|\phi_{\rm sim}|}{\sigma^2_\chi}}~.
\end{equation}

\section{Nuclear Star Cluster Timescales}
\label{app:timescales}
In this appendix, we examine and compare the relevant timescales for the orbital evolution of stellar-mass black holes bound to Sgr.~A$^\star$ and the propagation time of the ejecta. The latter is approximately, 
\begin{equation}
    t_{\rm prop} \simeq \frac{D_{\rm GC}}{v_\chi} \simeq 13 \, {\rm Myr} \left(\frac{D_{\rm GC}}{8.2 \ \rm kpc}\right)\left(\frac{600 \ \rm km/s}{v_\chi}\right)~,
\end{equation}
where we have normalized this expression to the lowest ejection velocity of interest, comparable to the conventional galactic escape cutoff.

We have approximated the black hole population in this region as being relaxed via two-body interactions. For reference, this timescale can be estimated in terms of the enclosed number of black holes and their average mass via \cite{Alexander:2005jz}
\begin{equation}
    t_{\rm rlx} \simeq \left(\frac{M_1}{M_2}\right)^2 \frac{P(r)}{2 \pi \log[0.4 \,N_{\rm BH}(<r)] N_{\rm BH}(<r)}~,
\end{equation}
where $M_1 \simeq 4.3 \times 10^6 \, M_\odot$ is Sgr.~A$^\star$'s mass, $M_2 \simeq 10 \, M_\odot$ is the assumed average mass of the orbiting black holes, $P(r)$ is the associated period for a circular orbit of radius $r$, and $N_{\rm BH}(<r)$ is the number of black holes enclosed within that corresponding volume. Utilizing Eq.~\eqref{eq:BH_dens_cusp} to determine $N_{\rm BH}$ contained within distance $r$, we find $t_{\rm rlx}$ to range between $(2 - 8) \times 10^8 \, \rm yr$ for the range of semimajor axes we consider. This is significantly shorter than the age of the nuclear star cluster, justifying our relaxation assumption. On the other hand, this timescale is considerably longer than the propagation time of the ejecta, allowing us to neglect any distortion of the spectrum between the emission and detection points.  

The relaxation time above describes the evolution of the semimajor axis distribution driven by distant two-body encounters. However, additional processes can evolve the eccentricity and orientation of the orbits. 
First, there is apsidal precession, $i.e.$ the secular advance of the pericenter. In our context, this is driven by the Newtonian potential of the extended mass around Sgr.~A$^\star$ rather than higher-order relativistic effects, and can be estimated as \cite{Alexander:2005jz} 
\begin{equation}
    t_{\rm aps} \simeq \frac{M_1}{M_2} \frac{P(r)}{N_{\rm BH}(<r)}~.
\end{equation}
Second, there exist resonant relaxation processes arising from the coherent torques exerted on a given orbit by the surrounding stellar-mass black holes. These are divided between scalar resonant relaxation, which redistributes the orbital eccentricities, and vector resonant relaxation, which randomizes the orientation of the orbital plane. Parametrically, these are shorter than the non-resonant relaxation time above \cite{2015MNRAS.448.3265K}, 
\begin{equation}
    t_{\rm rlx}^{\rm (sca)} \simeq \frac{M_1}{M_2} P(r)~,
\end{equation}
\begin{equation}
    t_{\rm rlx}^{\rm (vec)} \simeq \frac{M_1}{M_2} \frac{P(r)}{\sqrt{N_{\rm BH}(<r)}}~.
\end{equation}
Compared to the ejecta propagation time above, non-resonant two-body and resonant scalar relaxation operate on much longer timescales. Vector resonant relaxation and apsidal precession, by contrast, can be comparable to the propagation time. However, we emphasize that this process only induces secular changes on the orientation of the orbital plane, but leaves both the semimajor axis and eccentricity values unchanged. Since the ejection process, for weak deflections, can be approximated as isotropic ($cf.$ Eq.~\eqref{eq:iso_approx}), our results are independent of the orientation of the orbital plane, and therefore these timescales are largely irrelevant. Moreover, as we are integrating over a large number of orbiting black holes, we expect the effects of orbital plane re-orientations to average out in our estimates. 

One final process of potential relevance is the orbital decay due to gravitational wave emission. The average energy and angular momentum radiated over one orbital cycle are \cite{Landau:1975pou}
\begin{equation}
    \langle \dot{E}_{\rm grav} \rangle = - \frac{32 \, G^4 M^2_1 M^2_2 (M_1 + M_2)}{5 a^5 (1-e^2)^{7/2}} \left(1 + \frac{73}{24} \, e^2 + \frac{37}{96} \, e^4\right)~,
\end{equation}
\begin{equation}
    \langle \dot{J}_{\rm grav} \rangle = - \frac{32 \, G^{7/2} M^2_1 M^2_2 \sqrt{M_1 + M_2}}{5 a^{7/2} (1-e^2)^{2}} \left(1 + \frac{7}{8} \, e^2 \right)~.
\end{equation}
The associated timescales are
\begin{equation}
    t^{(E)}_{\rm grav} \simeq \frac{E_{\rm grav}}{\langle \dot{E}_{\rm grav} \rangle}~,
\end{equation}
\begin{equation}
    t^{(J)}_{\rm grav} \simeq \frac{J_{\rm grav}}{\langle \dot{J}_{\rm grav} \rangle}~,
\end{equation}
where $E_{\rm grav} = - G M_1 M_2/2a$ and $J_{\rm grav} \simeq M_2 \, \sqrt{G M_1 a (1-e^2)}$ respectively are the energy and angular momentum of the bound black hole.
Given the simulation grid shown in Fig.~\ref{fig:sim_grid_BW}, and assuming $M_2 = 10 \, M_\odot$ as before, we find the smallest $t^{(E)}_{\rm grav} \simeq 20 \, \rm Gyr$ and $t^{(J)}_{\rm grav} \simeq 409 \, \rm Gyr$ for the bins with $(a,e) \simeq (386 \, {\rm AU}, 0.94)$ and $(86 \, {\rm AU}, 0.41)$, respectively. As with non-resonant two-body relaxation, these timescales are generally long relative to the propagation time of the ejecta. We reach a qualitatively similar conclusion for the intermediate-mass black hole benchmark we consider in the main text. The secular evolution of the source due to gravitational radiation can therefore be neglected when computing the ejection spectrum at detection. 


\bibliographystyle{jhep}
\bibliography{biblio}

\end{document}